\documentclass{article}
\usepackage{blindtext}
\usepackage[a4paper, total={6.5in, 10.8in}]{geometry}
\usepackage{graphicx} % Required for inserting images
\usepackage{cite}
\usepackage{caption}
\usepackage{subcaption}
\usepackage{amsmath}
\usepackage{amssymb} 
\usepackage{amsthm}
\usepackage{soul}
\usepackage{placeins}
\usepackage{comment}
\usepackage{xcolor, float}
\usepackage{authblk}
\usepackage{hyperref}
\newcommand{\mycomment}[1]{}

\begin{document}

\title{Towards quantum computing for clinical trial design and optimization: A perspective on new opportunities and challenges}

\author[2]{Hakan Doga \footnote{Correspondence: \texttt{hakandoga@ibm.com}.}}
\author[7]{M. Emre Sahin}
\author[9]{Joao Bettencourt-Silva}
\author[4]{Anh Pham}
\author[5]{Eunyoung Kim}
\author[6]{Alan Andress}
\author[8]{Sudhir Saxena}
\author[11]{Aritra Bose}
\author[11]{Laxmi Parida}
\author[12;13]{Jan Lukas Robertus} 
\author[10]{Hideaki Kawaguchi} 
\author[8]{Radwa Soliman}
\author[1]{Daniel Blankenberg}

\affil[1]{The Cleveland Clinic, Lerner Research Institute, Cleveland, Ohio 44106, USA}
\affil[2]{IBM Quantum, Almaden Research Center, San Jose, California 95120, USA}
\affil[4]{Deloitte Consulting LLP, Atlanta, GA, USA}
\affil[5]{Deloitte Consulting LLP, Morristown, NJ, USA}
\affil[6]{Deloitte Consulting LLP, Philadelphia, PA, USA}
\affil[7]{The Hartree Centre, STFC, Sci-Tech Daresbury, Warrington, WA4 4AD, United Kingdom}
\affil[8]{My Next Health, Pleasanton, CA, USA}
\affil[9]{IBM Research Europe, Dublin, Ireland}
\affil[10]{Keio University, Quantum Computing Center, Yokohama, Japan}
\affil[11]{IBM Research, Yorktown Heights, NY}
\affil[12]{Imperial College London, London UK}
\affil[13]{Royal Brompton and Harefield Hospitals, Guy's and St Thomas' NHS Foundation Trust }
\date{}

\maketitle

%\tableofcontents{
%}
\begin{abstract}
  Clinical trials are pivotal in the drug discovery process to determine the safety and efficacy of a drug candidate. The high failure rates of these trials are attributed to deficiencies in clinical model development and protocol design. Improvements in the clinical drug design process could therefore yield significant benefits for all stakeholders involved. This paper examines the current challenges faced in clinical trial design and optimization, reviews established classical computational approaches, and introduces quantum algorithms aimed at enhancing these processes. Specifically, the focus is on three critical aspects: clinical trial simulations, site selection, and cohort identification. This study aims to provide a comprehensive framework that leverages quantum computing to innovate and refine the efficiency and effectiveness of clinical trials.
\end{abstract}

\section{Introduction} 

The clinical trial design is a process that undergoes significant rigor from many stakeholders and is ultimately aimed at proving both the safety and efficacy of a drug product across the target patient population that is impacted by the disease. By starting with clearly defined study objectives, a protocol can be developed which encompasses comprehensive criteria for how the trial should be executed while maximizing the safety of patients and the quality of data collected. Without proper selection and exclusion criteria for patients, the drug maker runs the risk of not completing the study or exhausting resources that could be better spent in other areas such as downstream patient support programs and lower drug prices.  

Patient cohort selection is one of the most important factors in the overall success of the trial such that an event rate is achieved that is statistically significant. The event can either represent the disease state or the treatment response or outcome. Adverse events and toxicity also must be tracked, and to be thorough must represent a diverse population to validate that the drug has evidence that demonstrates its safety and efficacy for all those it may be prescribed to for the intended use. Once the trial has design parameters it can be optimized for the patients, investigators, and sites that can be selected in order to balance the cost of the trial with a statistically significant outcome that supports a safe and effective claim that will be approved by the FDA. In order to evaluate patients, a set of multi-modal clinical, demographic, and geographic data is required to assess fit at an individual level. 

Quantum computing is a new frontier in advanced computing technologies that offers immense potential to enhance current classical computational methods. Based on the fundamentals of quantum mechanics \cite{dirac1981principles, heisenberg1973development}, quantum computing harnesses the power of entanglement, superposition and interference to manipulate quantum bits (qubits) and perform computations on quantum computers. Mapping the computational problem to a quantum computer, these principles allow quantum computers to perform certain computations potentially more efficiently. In particular, for certain classically intractable problems such as prime factorization \cite{shor1994algorithms}, unstructured database search \cite{grover1996fast}, solving linear systems \cite{harrow2009quantum}, computing topological invariants \cite{aharonov2006polynomial, lloyd2016quantum} and quantum simulations \cite{lloyd1996universal}, it has been proven theoretically that quantum computers provide speed ups in the computations.

\begin{figure}[h!]
    \centering
    \includegraphics[width=16.5cm]{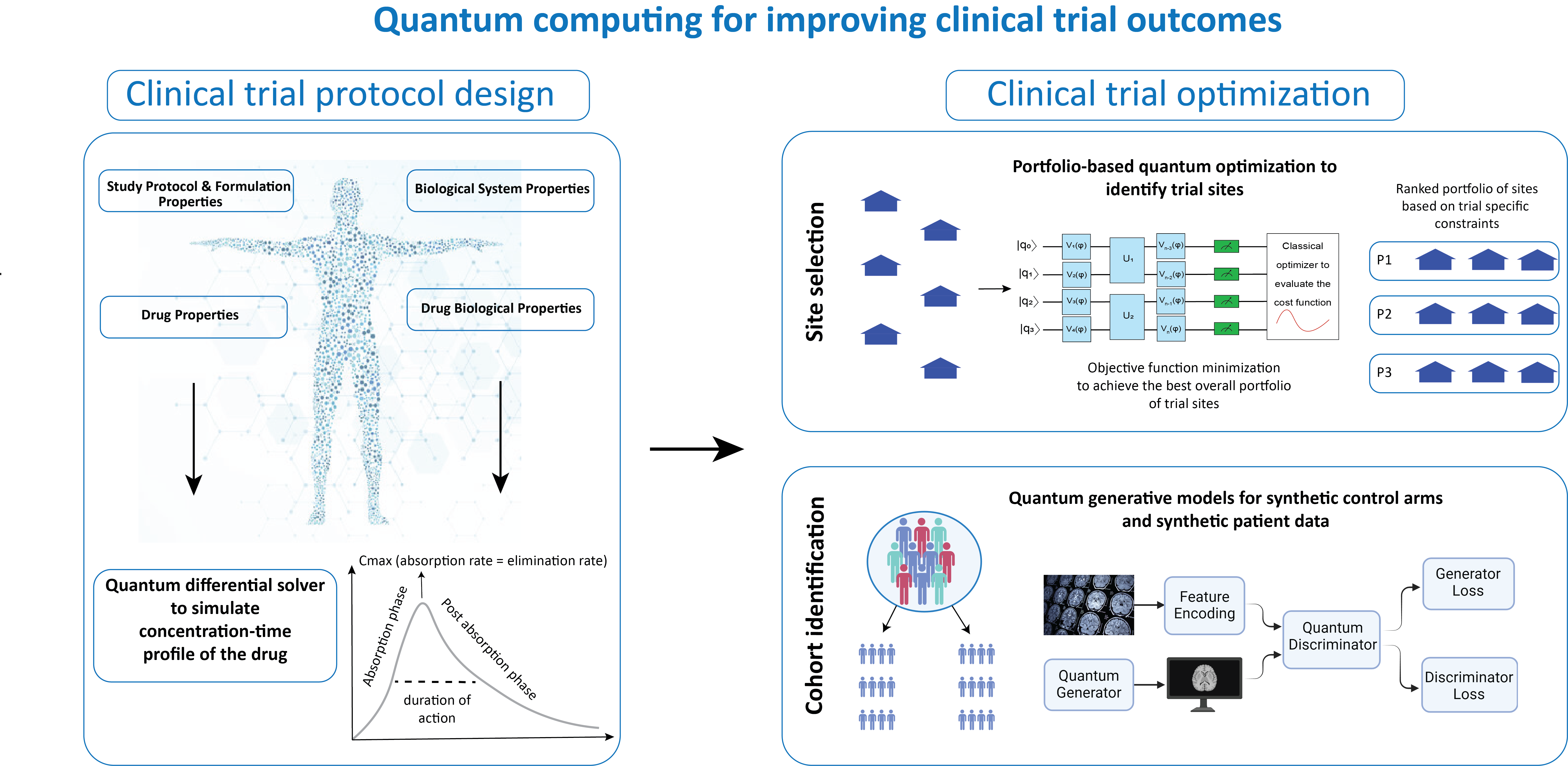}
    \caption{Protocol design is a crucial step for successfully optimizing the logistics of the trial. Using quantum differential solvers can enhance our understanding of mechanistic properties of the drug. Once protocols are set, this will inform the logistical optimization steps of the trial about how to select sites and how to identify the correct cohort. We propose portfolio-based quantum optimization to determine the best set of trial sites and quantum generative models to improve cohort identification.}
    \label{overall}
\end{figure}

Our work proposes using quantum computational methods to improve the outcome of clinical trials. To the best of our knowledge, this is the first comprehensive work that represents a first step on how to utilize quantum computing for clinical trial design and optimization. For different components of each subproblem we have identified, namely clinical trial simulation, cohort and site selection, we establish workflows with quantum optimization and quantum machine learning methods that can potentially improve the clinical trial design process (See Figure \ref{overall}). In Section \ref{challenges}, we describe the general challenges faced in today's clinical trial design and optimization methods. Section \ref{classical} discusses widely used classical computational methods and particular challenges that these methods have. Section \ref{quantum} provides a brief overview of quantum optimization and quantum machine learning methods, and delves deeper into the proposed quantum algorithms for enhancing clinical trial design. We conclude by discussing some future directions in the final section.

\section{Challenges in clinical trial design and optimization} \label{challenges}

% A survey of general challenges in clinical trial design

% 1 - Why drug candidates fail
Ninety-percent of drug candidates after they have entered clinical studies fail during trial phases I, II, III or during the final drug approval stage \cite{clinicalfailurerate}. Four main reasons for drug candidate failures in clinical trials have been reported: ``lack of clinical efficacy (40\%–50\%), unmanageable toxicity (30\%), poor drug-like properties (10\%–15\%), and lack of commercial needs and poor strategic planning (10\%)" \cite{clinicalfailurerate}. These are important factors that often require drug candidates to move back to the early development stages.

% 2 - Why clinical trials (design, conduct) fail 
As a small number of drug candidates are selected to advance to human trials, additional challenges emerge about trial design and conduct. During these stages, the most common reasons for clinical trial failures have been attributed to poor participant selection and recruitment. Since the early 2000s, different studies have reported recruitment-attributed fail rates of approximately 30\% \cite{Harrer_2019_AI_CTs, site_Njoku_2023} including reports that nearly 19\% of trial-initiated sites never recruited any participant \cite{site_Lamberti_2021}. In order to improve trial success rates, the FDA introduced new guidance specifically aimed at prospectively ``selecting a subset of the population in which the effect of the intervention can more readily be demonstrated" \cite{FDA_2019_Trial_Enrichment}. In parallel with cohort selection, site selection plays an important role for the success of the clinical trials. As reported in \cite{site_Fogel_2018}, choosing the optimal sites increase the chances of recruiting and retaining the target population. The location and easy access to trial site can prevent long travel times for patients, reducing potential patient drop outs. Site selection and cohort selection are two very important aspects of the clinical trial design process that needs to be modeled and planned symbiotically.  

% List of the challenges we plan to target

While these challenges may seem separate in nature, we believe there are underlying threads that connect them. Hence, we explore three overarching mechanisms that challenge the successful completion of trials
\begin{itemize}
    \item by constraining the study data (e.g. inadequate samples or cohort size, lack of historical site data),
    \item by introducing bias (e.g. imbalanced distributions, missing modalities in data),
    \item or by detecting  study endpoints or outcomes related to clinical effectiveness (e.g. toxicity-efficacy, optimal dosage and frequency).
\end{itemize}  

For each of these mechanisms there is a range of contributing factors that may be cohort-specific (e.g. patient identification, insufficient participants),  operational (e.g. inadequate site selection, resources and costs) or a combination of both. In this section and the following sections, we expand and detail these mechanisms and their impact on trial outcomes. These factors may arise during different stages of trial design and conduct, which we address in this work in three areas: clinical trial simulations, site selection and cohort identification.

% is Simulation supposed to tackle the issues FDA addresses in \cite{FDA_2019_Trial_Enrichment}? Does this also include synthetic control arms?

\subsection{Clinical trial simulations}

The journey of drug development is notably long and expensive, often spanning 10 years and necessitating an investment of over \$1-2 billion, with the rate of success lingering below 7\%, due to the complexity of the pre-clinical and clinical trials essential for validating safety and efficacy \cite{dimasi2016innovation, wong2019estimation, wouters2020estimated}. Factors contributing to this scenario include the inadequate consideration of biophysical and physiological characteristics of molecules in initial protocol definitions, leading to disparities between in vitro experiments and in vivo studies \cite{otava2015identification, gonzalez2015vitro}, and between disease models in animals and humans \cite{kim2020human}. Furthermore, the preliminary stage calculations for dosage and the mechanism of action often fall short in accurately foreseeing efficacy and toxicity in humans \cite{kim2020human}, compounded by less than optimal event rates and high dropout rates in clinical trials. To mitigate these challenges, it is important to predict the effects of drugs on specific subpopulations, segmented by genomic and epigenetic profiles \cite{hamburg2010path}. The development of sophisticated models that can accurately predict efficacy and toxicity, particularly by evaluating the mechanism of action in human pharmacokinetics and pharmacodynamics (PBPK/PD), is important. A precise determination of drug dosages and regimens that take into account the full spectrum of human physiology and potential interactions with other drugs is crucial for enhancing the drug development paradigm.

Optimizing clinical trials presently leans more towards an empirical approach rather than a rigorous scientific method. Establishing a robust scientific foundation for the design and optimization of the protocol, along with cohort/site selection, requires the identification and collection of sufficient data points. These data points should provide insights necessary for informed protocol design.

\begin{enumerate}
\item   Understanding pharmacokinetics, pharmacodynamics and pharmacogenomics – Designing a clinical trial with real people for a new molecule requires a thorough understanding of how a body would process that drug, and how the drug would impact the body.
There is, thus, an underlying requirement to consider the pharmacokinetics (what happens to the molecule as it passes through the human body) and pharmacodynamics (what happens to the body as the molecule passes through it) properties related to the molecule in question for the condition(s) in question. Furthermore, understanding the  genomic and epigenetic influence on underlying patient physiology based properties, and their interplay with the drug molecule characteristics is critical, as this serves as the basis for designing the trial protocol and the criteria for site and cohort selection.

\item   Understanding dosage and mechanism of action - without proper prediction algorithms to identify the correct dosage and the mechanism of action at a preclinical stage, it is not viable to construct a true predictor of the efficacy toxicity in real humans. Preclinical efficacy data is not an accurate predictor of efficacy in clinical trials. There are biological discrepancies  between in vitro and in vivo, and  between disease in animal and humans. Consequently designing clinical trial protocols based on in vitro and in vivo data can result in lower event rates, and higher drop out rates.

\end{enumerate}

\mycomment{\textcolor{red}{To effectively model and simulate clinical trials in this increasingly digital landscape, it is crucial to consider the implications of data management practices. The transition from physical to virtual storage of patient data by medical institutions has proven to be highly advantageous, allowing for remote access and updates while alleviating concerns regarding physical storage limitations for paper filing systems. However, the virtual storage of such sensitive information makes medical institutions a prime target for cyber security threats. Recent reports indicate that up to 94\% of medical institutions have fallen victim to cyber-attacks \cite{perakslis2014cybersecurityinhealthcare}, highlighting the necessity for robust cybersecurity measures to protect confidential patient data. The increasing reliance on software-dependent technology within the medical domain increases the possible fallout of cybersecurity breaches. As such, each technological advancement can introduce new vulnerabilities. The consequences range from compromised personal privacy \cite{cyberbreaches} to life-threatening scenarios, such as remote access to vital medical devices like pacemakers \cite{kramer2017cybersecurity}. Despite hundreds of incidents and millions in payouts from medical bodies, cybersecurity considerations remain largely overlooked, with just 2.13\% of medical-related software products acknowledging its importance \cite{Sterne025374}.}}

\subsection{Site Selection} 

%(1) Importance
Adequate site selection is one of the most critical steps for the successful completion of a trial \cite{site_Bieske_2023}. An appropriate trial site must be equipped with necessary study-specific infrastructure and resources, such as staff and equipment, as well as provide access to a catchment area that enables the selection and recruitment of an adequate number of participants. Trials may also be conducted across multiple locations to ensure that a more representative sample of the population can be studied and, in some cases, expanding to additional sites may also help counteract trial delays and bottlenecks \cite{site_Bieske_2023}. 

%(2) Site selection and trial failures and Features associated with success 
Site-specific characteristics are therefore important and can contribute to serious downstream bottlenecks for the timely conduct and completion of trials. Examples of cohort-specific characteristics include a non-homogeneous, diverse population within a site's catchment area, and operational-specific characteristics include quality electronic records and available data capture infrastructure, domain-specific expert trialists, equipment or laboratories.

%(3) How are trialists selecting sites today
The majority of industry experts select a trial site based on three dimensions \cite{site_Bieske_2023}: (1) Knowledge about the site (e.g. experience in conducting trials, site success rates, therapeutic specialization); (2) Patient access (e.g. recruitment performance or access to required patient population); and (3) Working culture (e.g. known investigators' interest and commitment to driving studies). 

%(4) How should we be selecting moving forward
Site selection should also take into account  geographical considerations that are specific to the site location such as an increased incidence of certain diseases, particular exposures to environmental factors, possible burdens for patient participation and retention, or regulatory requirements. Overall, existing approaches and algorithms to rank the most adequate sites for a trial require modelling these complex and diverse criteria involving both cohort-specific as well as operational-specific characteristics. Successful methods must cope with limited data availability and site-specific biases, which further make comparisons across sites a non-trivial challenge.

%The altitude example for sickle cell and other hematological disorders ... higher alt -> higher production of red blood cells, higher hematocrit -> CVD/stroke/vascular events

% 

\subsection{Cohort Identification} 

Identifying the relevant inclusion and exclusion criteria so that patient selection can occur adequately at different clinical sites is a challenging task. For instance, if the eligibility criteria are too strict, even though they can satisfy clinical concerns, this can create barriers to trial enrolment and recruitment \cite{Strict_criteria_oncology,10.1001/jama.297.11.1233}. In addition, when the criteria are narrowly defined, they can prevent trial results from being generalized\cite{Generalization_criteria}. Furthermore, trials do not often include certain population groups based on age \cite{10.1001/jama.2019.17016}, gender \cite{10.6004/jnccn.2017.7050}, and ethnicity\cite{doi:10.1177/107327481602300404}. As a result, new regulation from the US Food and Drug Administration (FDA) requires adequate representation of the population in clinical trials \cite{FDA_guideline}, which requires the modification of the exclusion and inclusion criteria and protocols to balance clinical concern with adequate sample size and representation. However, currently there is a lack of consensus on how to remove specific criteria \cite{doi:10.1200/JCO.2015.62.1854}.

There are noticeable race, ethnicity and gender disparities in populations participating in clinical trials. This causes issues such as the lack of generalization of data emanating from trials and lack of information on the effectiveness of drugs in certain population groups~\cite{yekeduz2021assessing}. In many cases, the implementation of current guidelines on clinical trial design is sub-optimal in the population at large, but especially among ethnic minority groups, women, people living in rural communities, uninsured or under-insured individuals, lower education and income, and in general individuals in vulnerable socio-economical conditions~\cite{khoury2022health}. Clinical trials lack diversity due to  historical limitations, but also due to their early geo-spatially restricted origins~\cite{zaaijer2021ancestry} as well as the multiple barriers to participation which disproportionately affect certain population groups more than others \cite{RODRIGUEZTORRES2021100829}. Thus, equity is an important factor in clinical trial design to improve generalizability, misclassification of drugs, targeted personalized treatment, and reduce implicit bias. 

To further enhance the personalization of treatment, reducing disparities, and lowering the number of enrolled required patients, enrichment methods like prognostic and predictive strategies have been proposed to overcome several deficiencies in current cohort identification and recruitment processes. Prognostic enrichment refers to the targeted recruitment of patients with certain diseases outcome and progression. Alternatively, another process known as predictive enrichment refers to the recruitment of patient population with the potential for maximum positive response to the trialed therapy or drugs. Such a method will help to reduce the number of patients required for the intervention group, while prognostic enrichment will help to enhance the power of the study. In practice, these two methods have been used as complements to each other since they can benefit trials for personalized medicine treatment where a prediction of treatment's efficacy is intricately linked to the disease progression like sepsis \cite{Stanski2020, Wong2016-ao}. However, the task of predicting disease progression, as well as treatment outcome using using patients' genome-to-exposome profile \cite{10.1117/12.2178956} is a scientifically challenging task. 

From a technical perspective, the cohort identification process has relied on expert knowledge to select the criteria based on physicians' clinical expertise \cite{10.1136/amiajnl-2013-001935} in conjunction with manual queries in patients' medical records. However, this can potentially result in bias in identifying the eligibility criteria \cite{10.1093/biostatistics/kxn010}, as well as slowing down the recruitment process. As a result, many computational methods like AI/machine learning (ML) models have been proposed to enhance the criteria selection process, to complement the clinical expertise and experience of the trial investigator \cite{spasic2019cohort}. Despite major benefits that have been demonstrated by applying ML models to cohort selection, many challenges still remain. Most prominently is the heterogeneous nature of biomedical and healthcare datasets. Specifically, electronic health record (EHR) are known to have several data quality issues from medical codes, sparse or missing data, as well as the differences in how medical data is recorded when comparing records between sites \cite{Wells2013StrategiesFH}. In addition, new personalized medical treatments targeting specific biomarkers will also require other data types in conjunction with EHRs like omics data \cite{Rockowitz2020}, making the application of data driven techniques, such as machine learning, challenging due to the heterogeneous nature of different data types \cite{Kline2022}. Furthermore, for rare disease cases, it can be difficult to obtain datasets with the appropriate sample sizes, thus making it difficult to train robust machine learning models \cite{Hee2017}. An emerging approach to tackle sample size limitations is the use of uncontrolled trials where synthetic control arms (SCA) may be used. Unlike traditional RCTs, in uncontrolled trials, only the treatment arm needs to be recruited and the controls may be generated using existing data (e.g., EHRs) or synthetically generated data based on the trial protocol. ML can support this task by automatically generating synthetic control arms to supplement single arm treatment-only trials. However, this approach may have unique challenges with regards to generalizability, interpretation, and statistical methodology \cite{synthetic_arms_thorlund2020}.

Finally, a potential concern for the application of machine learning techniques in clinical trials is the lack of explainability due to their black-box nature. For instance, ML models have been proven to be effective to predict the personalized advantage index (PAI), which is a measurement of the optimal treatment within a patient cohort \cite{ML_PAI} with psychiatric conditions. However, the potential drawback of using machine learning to predict PAI values is the lack of exaplainability in the decision making process of the model. As a result, it is imperative that explainability should be built into the model to inform the decision outcome \cite{Mellem2021}, thus satisfying regulatory concern as well as enabling cross validation/confirmation by clinicians conducting the trials. 

\section{Classical computational methods in clinical trial design and optimization} \label{classical}

\subsection{Clinical trial simulations}

In computational clinical trials, the key challenge is to effectively simulate the outcomes of clinical trials using computational models  \cite{gutierrez2021methods}. The goal is to predict drug efficacy, side effects, and outcomes without conducting physical trials, which has the potential to lower costs, save time, and address ethical concerns inherent in conventional trials \cite{inan2020digitizing}. This process utilizes a range of techniques, including advanced algorithms \cite{hekler2016advancing}, statistical methods \cite{liao2016sample}, machine learning, and various simulation approaches such as patient virtualization and pharmacokinetic/pharmacodynamic (PK/PD) modeling. The focus is on virtual rather than physical simulations, to accurately forecast outcomes as they would occur in the real world.

%The term "computational complexity" in this context refers to the computational resources required for these simulations, which are determined by factors such as the complexity of the models, the size of the simulated populations, the intricacies of the diseases being modeled, and the level of detail in the patient data utilized.

The modeling of pharmacokinetics and pharmacodynamics plays an important role in evaluating drug safety and efficacy, confronting the challenges posed by biological complexity and the extensive data required. As these trials progress, there is a need to enhance the accuracy and efficiency of these models. Overcoming these obstacles requires advancements in computational methodologies and robust datasets.

PBPK is a powerful modeling technique that requires computational tools to solve ordinary differential equations. These equations describes the movement of the drug between the different tissues/organs in the biological systems. PBPK modeling represents a mechanistic approach that integrates 2 important components : (1) drug-dependent component that includes properties specific to the drug and (2) drug-independent component that includes biological system properties, study protocol and formulation properties. A third component related to both the drug and the biological system, allows for a priori simulation of drug concentration–time profiles.  Such predictions are of high pharmacological value because they enable estimating drug exposure not only in plasma but also at the site of action, which may be difficult or impossible to measure experimentally  \cite{kuepfer2016applied}.

PBPK models simulate the concentration–time profile of a drug in a species by integrating the physicochemical properties of the compound with the physiology of the species. Being mechanistic, PBPK models can be used to simulate and to predict the next set of data and to plan the next experiment.

Simulated clinical trials using PBPK population modeling of these complex clinical scenarios need to be fully explored. \cite{hartmanshenn2016physiologically}.
Many reviews emphasized the significance of PBPK modeling for personalized medicine. \cite{rowland2011physiologically, chetty2014applications, jamei2009population}. These models have the potential to guide physicians in prescribing the right dosing regimen \cite{hamburg2010path, huang2013utility} as these models utilize the relationship between the drug dose, physiology and plasma concentration, which helps determine dosing regimen within a target therapeutic window \cite{kang2009overview}. Such models can be used to design more inclusive clinical trials, by conducting in silico simulations across diverse population groups to evaluate patient risk and benefits, and providing dosage guidance \cite{strougo2011semiphysiological}.

PK/PD modeling can optimize the design of clinical trials, guide the dose and regimen that should be tested further, help evaluate proof of mechanism in humans, anticipate the effect in certain subpopulations, and better predict drug-drug interactions; all of these effects could lead to a more efficient drug development process. Because of certain peculiarities of immunotherapies, such as PK and PD characteristics, PK/PD modeling could be particularly relevant and thus have an important impact on decision making during the development of these agents. 

By looking at the disease behavior patterns, we can actually classify the drug candidates based on their potency and based on the tissue exposure. One can also predict drug - drug interaction, and use that to determine the dosage and frequency regimen before initiating trial design.

\subsection{Site Selection}

Clinical trial site selection is a complex process that requires both quantitative and qualitative analysis of multiple factors \cite{gehring2013factors}. While the focus of this work is on adequate site selection for clinical trials, similar problems occur when selecting optimal sites in other industries such as renewable energy sites \cite{shao2020review}, landfill sites \cite{csener2006landfill} or facility sites \cite{liang1991fuzzy}, and we observe similar computational methods employed to analyze the relevant factors for optimal site selection. In the case of clinical trials, some of these factors can be more generic while the other are trial-specific. Depending on the scope, the overall goals and the logistics of the particular trial, sites are selected to maximize a successful outcome. Viewed this way, site selection can be seen as an optimization problem over multiple parameters or a machine learning problem where we would like to understand most significant features to train a model for predicting best sites.

When selecting appropriate sites for clinical trials, most classical methods rely on historical trial data from the candidate sites and their catchment population. Site-specific data may be collected, for example, from publicly available databases such as ClinicalTrials.gov, through Contract Research Organizations (CROs), or from proprietary data sets. Cohort or patient-specific information may be collected from electronic health records (EHRs), databases of geographical distribution statistics of disease and patient demographic characteristics. When available, this information can be used by trialists to rank trial sites that are more likely to be successful in recruiting participants and completing the trial.  

Ranking based approaches may use statistical summaries \cite{site_portfolio_opt, theodorou2023framm} and portfolio optimization \cite{jorion1992portfolio}. The latter is a constrained utility-maximization problem where an optimal portfolio (i.e. a trial site and its characteristics) is selected from a pool of considered portfolios (i.e. candidate trial sites) according to an objective (i.e. successful trial completion or successful trial participant recruitment). 

Additionally, machine learning based approaches have also been used to select ideal trial sites. Traditionally, site selection has been conducted by experts with heuristics or rules \cite{Hurtado-Chonge014796}. However, rule-based methods are extremely costly, time-consuming and dependent on existing trialists' expert knowledge. More recently, methods to predict which sites should be selected and to match sites with clinical trials by efficiently using machine learning for site selection have been attracting attention \cite{biswal2019doctor2vec, 10.1093/jamia/ocz064}. For example, a method that introduces constraints to ensure the diversity of patients to be enrolled in a ranking-based method \cite{srinivasa2022clinical} has been developed and can be easily implemented \cite{wang2023pytrial}. 

Knowledge graph based methods have been used to create novel representations and perform analyses using embeddings and graph learning models \cite{site_CTKG, cui2023survey}. Although these are promising approaches, both graph-based and non-graph-based neural networks require large amounts of training data to achieve model accuracy performance. However, site-specific and patient-specific data are heterogeneous, sparse, and are often missing, which further add obstacles to current classical methods.

%Adaptive Trial optimization could also be added here

\subsection{Cohort Identification} 
Eligibility criteria are identified to select patients for clinical trials by the investigators at the trial sites. Currently there are different techniques which have been used to recruit participants for trials. Specifically, the traditional methods of cohort selections have been rule-based methods, which can be time intensive. Furthermore, longer cohort selection process can negatively impact on the timeliness of the clinical trial. Consequently, researchers have taken advantage of data-driven techniques like machine learning algorithms to speed up the screening of participants in clinical trials based on inclusion-exclusion criteria.  In this section, we will review exiting statistical, optimization, and machine learning methods which have been used to screen patients based on eligibility criteria. In addition, more advanced applications of machine learning to identify relevant subgroups for personalized medicine treatments, as well as the emergence of generative models to generate synthetic data will also be reviewed.
\subsubsection{Rule based approach}
The traditional methods of cohort selections have used rule-base methods. Those rules are often crafted to satisfy specific criteria for a clinical trial or specific types of trials. Longer cohort selection time makes negative impact on the timeliness of the clinical trial. Crafting those rules takes a lot of time and efforts including medical experts, yet it has less relevance to future clinical trials cohort selection. Also, although candidates are suitable to clinical trial criteria, it is harder to predict their length of stay (LOS) or discover risk factors of LOS in advance using the rule based approach. The candidates’ prolonged LOS and the finding the risk factors of LOS are key factors for successful cohort selection. 

\subsubsection{Machine learning} 
Using machine learning algorithms, researchers can not only classify, cluster, predict more suitable candidates but also have ability to predict their length of stay or discover risk factors or group for LOS. Those were hidden factors that may have not been defined as a rule. Many studies that utilized machine learning methods have been tested on electronic health records (EHRs). However due to the complexity of EHR dataset, multiple ML models need to be tested to maximize the prediction outcome. For example, Segura-Bedma et al. \cite{segura2019cohort} applied different deep learning methods to select robust cohort selection. Similarly, Chen et al. \cite{chen2023machine} used four machine learning models – Neural Network (NN), random forest, histogram-based gradient boosting, and k-nearest neighbor to identify patients at high risk for prolonged LOS following primary total hip arthroplasty. Other researchers leveraged NLP (Natural Language Processing) and used bag of words document representation model, which considers the text as an unordered set of words, for 13 classifiers and other pattern matching approach. To select the candidate, the authors investigated four different classification methods - support vector machine, logistic regression, naïve Bayesian classifier, and gradient tree boosting\cite{krishnasamy2014hybrid}. More advanced development in ML model like transformer-based approach designed specifically for EHR \cite{li2020behrt} has shown promising results in performance with potential for generalization. 

%Another prominent application of ML models is predicting the outcome of specific treatment within certain population groups due to biological or physiological biomarkers using genome-to-exposome patients' profile \cite{10.1117/12.2178956}. For instance, ensemble machine learning model has been used to predict the treatment outcome of patients with type 2 diabetes given DPP-4 inhibitors in comparison to conventional therapy \cite{Berchialla2022}. Such an approach can benefit the predictive enrichment strategy in clinical trial, whereby certain population with the maximum potential of positive response to the therapy is targeted for recruitment to the intervention group, with the goal of minimizing the failure rate. In addition, ML can also play an important role in selecting patients with a higher disease-related outcome of interests, e.g. focusing on recruitment of patients with high mortality rate \cite{https://doi.org/10.1002/ejhf.2528} for certain disease. %

In addition to typical ML models and advanced transformer model, generative models like the generative adversarial network (GAN)\cite{goodfellow2020generative} has also shown promising applications to generate synthetic data \cite{KRENMAYR2022101118, platt2024creating}. In clinical trials, synthetic data can be used to create synthetic patients, with statistical properties of real patients but are not linked them, thus maintaining patients' privacy and confidentiality. These synthetic patients can also act as a synthetic control arm, in conjunction with the real patients in the intervention group. As a result, the usage of the synthetic control group can reduce the number of participants needed to be recruited in a trial, thus speeding up the trial process and reducing the budget requirement.  

GAN structure consists of a generator and a discriminator \cite{10.1145/3422622}. The generator's role is to generate synthetic data, while the discriminator plays the role of a classifier to distinguish between fake and real data. These two models compete against each other, thus resulting in a Nash equilibrium whereby the quality of the synthetic data produced by the generator can resemble the real data. However, GAN requires the usage of pseudo-random number generator to inject noise to the generator, thus potentially resulting in large computation time, and potential statistical bias in the model. 

%Current research direction is using Electronic Health Record (EHR) dataset and machine learning method. EHR has many informative information as well as medical history. The work \cite{liu2023toward} tried to find a Universal Cohort Representation Learning Framing for Electronic Health Record Analysis. EHR has many informative information as well as medical history. EHR is used in both rule-based and ML approaches%.

Another challenge for current ML techniques is how to embed and train models that can process multi-model data to improve the selection of eligibility criteria. The multi-modal dataset typically involves different data formats such as EHR, which typically represented as tabular data (e.g., age, demographics), image data (e.g., x-rays, magnetic resonance imaging, pathology slides), time-series data (e.g., pulse oximetry, ultra-sounds, wearable sensors), structured sequence data (e.g., genomics, proteomics, metabo-lomics) and unstructured sequence data (e.g., notes, forms, written reports, video) among other sources \cite{Soenksen2022}.  Fusion strategies have been proposed as a solution to ingest and train machine learning models for multi-model dataset, depending on the stage or methods of the data fusion, it is called early fusion or late fusion or joint fusion \cite{Kline2022, mohsen2022artificial}. This includes combining features from the relevant data, extracting the relevant features from dataset and aggregating them, or applying assemble learning models for different data modalities \cite{Soenksen2022}.

\subsubsection{Personalized treatment, genomics, and equity}
Cohort selection is not just restricted to rule-based and machine learning aspects. Trial participants are also chosen for the trial based on their underlying similarities stemming from genomics. In some cases, patients might have the same tissue-of-origin diagnosis, or genomic alterations. In recent years, due to advancement in genomics technology and subsequently, better understanding of the biological underpinnings of complex diseases, there is a consensus that complex diseases such as cancer, cardiovascular disease, neurological diseases need customization to an individual for optimal therapy. Hence, next-generation trials need to be patient-centered rather than drug-centered~\cite{fountzilas2022clinical}.

Trials modeled around specific biomarkers for certain drugs treating heterogeneous complex diseases aim to match patients with specific interventions based on their molecular characteristics. Single and multiple integral biomarker trials classified as basket, umbrella, and adaptive trials have been proposed as novel approaches for clinical trial design for personalized treatment. Basket trials group patients with same molecular abnormality regardless of disease outcome; umbrella trial design tests multiple drugs targeting a common pathway, but in genetically distinct subgroups; while adaptive trails allow for adjustments to the design mid-course based on feedback from the targeted cohort~\cite{hu2019biomarker}. These biomarker-driven approaches also necessitate understanding of genetic diversity with respect to drug target identification. Machine learning models and natural language processing techniques as described above can help identify individuals with a high likelihood of responding positively to a specific treatment, as well as survey social determinants of health that could impact trial participation and outcomes. For instance, ensemble machine learning model has been used to predict the treatment outcome of patients with type 2 diabetes given DPP-4 inhibitors in comparison to conventional therapy \cite{Berchialla2022}.

Extending specific biomarker oriented trials to increase clinical trial efficiency, another way is to selectively enroll participants based on clinical and biological characteristics~\cite{food2019drug}. Lately, polygenic risk scores (PRS) which take into effect the cumulative impact of many genetic variants across the genome to compute potential genetic predisposition or risk of a disease have been considered for cohort identification~\cite{fahed2022potential}. It was observed in post hoc analyses of clinical trials that healthy individuals with highest PRS for cardiovascular disease demonstrated the greatest benefit when randomized to statin or placebo to prevent cardiovascular disease, with a 44\% relative risk reduction versus 24\% in the remainder of participants~\cite{natarajan2017polygenic, mega2015genetic}. Beyond post hoc analyses, PRS can also be used to understand the presence of racial disparities in large biobanks which are used to identify cohorts in clinical trials~\cite{bose2023role}. However, PRS are known to suffer from lack of generalizability and loss of predictive power in non-European populations due to the predominant focus of genome-wide association studies on European populations~\cite{martin2019clinical}. Recent studies have shown that even with a small number of non-European participants predictive power of PRS increases in underrepresented populations~\cite{wang2023global}. Machine learning methods are often applied to improve generalizability of PRS across diverse cohorts for complex diseases~\cite{ge2019polygenic, reyes2023fairprs, platt2024ai}.

By incorporating machine learning algorithms into response-adaptive randomization designs, trials can achieve better performance than traditional designs, resulting in more ethical and appropriate clinical trials~\cite{wang2022application}.

\section{Quantum computational methods in clinical trial design and optimization} \label{quantum}

In this section, we briefly discuss some quantum algorithms that are relevant for this work and present our perspective on how to utilize quantum computational methods for enhancing clinical trial design and optimization. For a more in-depth discussion of quantum computing, especially in the context of healthcare and life sciences, the reader is referred to Sections 1 and 2 of \cite{basu2023quantumenabled}.

\subsection{Quantum Optimization and Quantum Machine Learning Algorithms}

In general, optimization problems aim to find the global minima/maxima of an objective function. As the parameter space for the problem increases, searching for this global extremum point becomes harder and more costly for classical methods. Quantum optimization methods \cite{ barkoutsos2020improving_QO, moll2018quantum_QO_var, symons2023practitioner} have been proposed, studied and implemented with the goal of achieving a speed-up over the classical optimization techniques. For near-term implementations of quantum optimization algorithms, a common method is using variational quantum algorithms \cite{cerezo2021variational}. These are quantum-classical hybrid algorithms that can efficiently explore the high-dimensional Hilbert space to find the optimal solution. To that end, a trial quantum state is prepared and appropriate mapping of the problem Hamiltonian is set up. Classical optimizer updates the quantum state in every iteration of this pipeline until a solution state is recovered. The work \cite{abbas2023quantum_optimization} provides a rigorous and detailed analysis of the current landscape for quantum optimization algorithms, their complexity and potential for certain optimization problems such as maximum independent set (MIS), knapsack/market share problem, sports timetabling problems and several others.

Quantum machine learning (QML) \cite{dalzell2023quantum_ML, schuld2015introduction_QML, biamonte2017quantum_ML, cerezo2022challenges_QML} is another growing field within quantum computing with the potential to improve data analysis. In a typical machine learning scheme, an algorithm is designed to learn the structure of the data, and identify patterns for various prediction and inference tasks. These learning algorithms are usually considered under three main categories, supervised machine learning, unsupervised machine learning and reinforcement learning. While data is labeled in supervised ML for the training of the model, in unsupervised learning the algorithm finds the patterns in the data and learns them. In reinforcement learning, the training is done on a reward/punishment basis. Despite there being no general and complete QML theory, all these machine learning schemes have counterparts in QML. For example, quantum neural networks (QNN) \cite{cong2019quantum_CNN} can be used for computer vision tasks for labeled data or solve a MAXCUT problem on a graph by learning the distances as an unsupervised learning model \cite{otterbach2017unsupervised}. A more in-depth look into the field of QML can be found in \cite{cerezo2022challenges_QML} with a main focus on QNNs and quantum kernel methods.

\begin{figure}[h!]
    \centering
    \includegraphics[width=17cm]{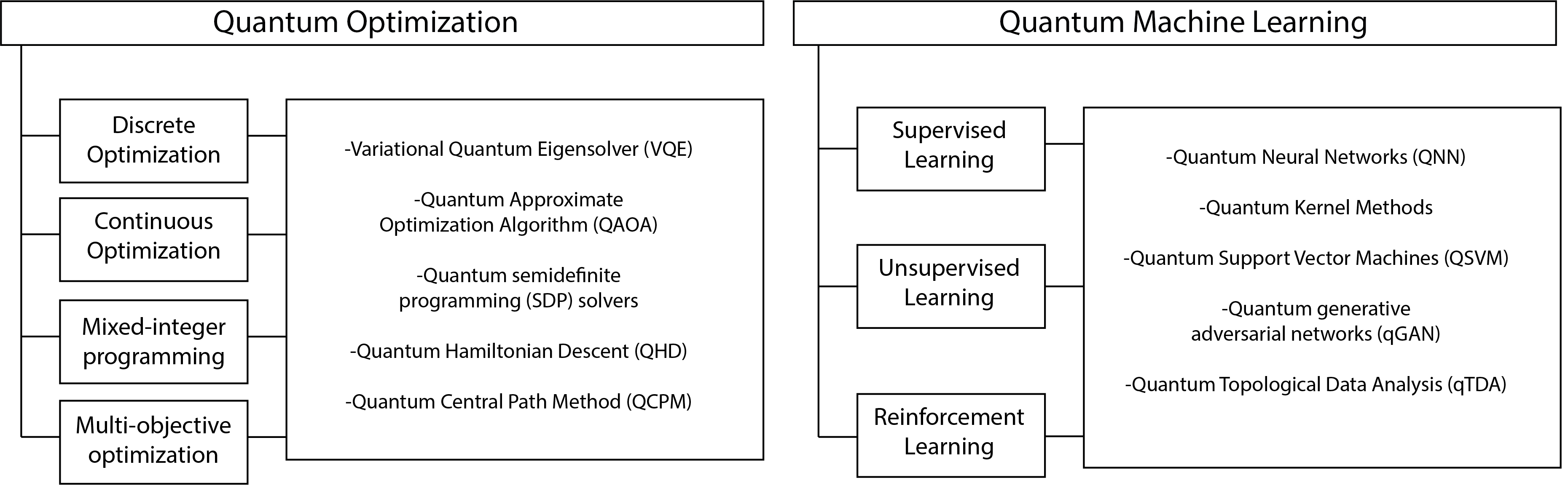}
    \caption{Table lists some of the problem classes within optimization and machine learning problems, along with some commonly used examples of quantum algorithms for optimization and machine learning.}
    \label{quantum-algo}
\end{figure}

In general, it is an ongoing and active research area to map the quantum algorithms to various use cases and identify how and where quantum computing can improve the existing computational methods. This interest has led quantum computing research to find a large variety of applications in various fields ranging from molecular chemistry and molecular simulation \cite{atzori2019second, kovyrshin2023nonadiabatic, nykanen2023toward, kovyrshin2023quantum, kandala2017hardware, rice2021quantum, eddins2022doubling, barison2022quantum, motta2023quantum, liepuoniute2024simulation, rossmannek2023quantum}, material sciences \cite{bauer2020quantum}, encryption \cite{ducas2018crystals, scholten2024assessing}, search and optimization \cite{glover2018tutorial, gambella2020multiblock}, protein structure prediction and drug discovery \cite{mensa2023quantum, doga2023perspective}, cell-centric therapeutics \cite{basu2023quantumenabled}, financial optimization \cite{orus2019quantum}, computational pathology \cite{sahin2024efficient}, causal discovery \cite{10.1371/journal.pone.0283933}, and other potential applications in health and medicine \cite{flother2023state}.

% For the case of Mixed-integer optimization consider "Multiblock ADMM Heuristics for Mixed-Binary Optimization on Classical and Quantum Computers", CLAUDIO GAMBELLA, ANDREA SIMONETTO.

\subsection{Clinical trial simulations}

 The limitation of the PBPK models originate from the uncertainty in regard to the degree of accuracy and reliability of its input. The reduced confidence in input data is understandable since  accuracy of PBPK models comes from the quality of its input parameters. Therefore, if the incorporated data contain errors, inconsistencies or miscalculations, the resulting simulations may not truly represent in vivo conditions and performance. Inaccuracies in input data may originate from a variety of reasons such as the high inherent variability of biological systems \cite{johansson2008physiologically, Nestorov2003}
 Another challenge of PBPK models is severe lack of detailed databases for
physiological parameters, particularly for certain subpopulations. Consequently, much of the input data must be obtained from multiple sources.  \cite{khalil2011physiologically}
Finally, a deficiency of information regarding human physiology/anatomy and information related to drug metabolism and active transport leads to poor prediction of the observed data \cite{johansson2008physiologically}. Clinical trials necessitate the use of high-quality data \cite{krishnankutty2012data, subbiah2023next}, typically characterized by a limited dataset size. QML algorithms have demonstrated effectiveness in scenarios with few data, as evidenced by various studies \cite{caro2022generalization, gil2024understanding}. To bridge the gap between the necessity for high-quality data in clinical trials and the demonstrated effectiveness QML algorithms in scenarios with limited datasets, it is essential to find efficient embeddings that can accurately translate classical data characteristics into quantum-compatible formats. Specific metrics for analyzing the classical data and its corresponding feature map can be adopted \cite{gil2024understanding,caro2021encoding, banchi2021generalization, huang2021power, abbas2021power, li2022concentration}, allowing for the potential of QML algorithms to be more effectively realized for these specialized applications.

 A mechanistic understanding of PK processes and a deeper knowledge of these specialized physiologies is key for improving confidence in specialized PBPK models and therefore achieving the ultimate goal of personalized medicine \cite{jones2015physiologically}. Integrating systems biology (target expression) with PBPK and Pharmacodynamics modeling will further increase the usefulness and adoption of PBPK modeling. This is even more relevant in personalized medicine where individual patient outcomes can be improved by delivering the personalized optimal dose at the right time \cite{ideker2001new, claudino2007metabolomics, peters2021physiologically}. The use of imaging techniques for tissue partitioning to eliminate the uncertainty related to predicted tissue-to unbound plasma partition coefficients (Kp), key set of parameters used to characterize the distribution or “movement” of drugs into different tissues in the body \cite{jones2013basic}. The development of more stable and viable in vitro systems is also important\cite{maguire2009design}, as is development of competitive binding assay for highly bound compounds \cite{schuhmacher2004high}.

\begin{figure}[h!]
    \centering
    \includegraphics[width=16.5cm]{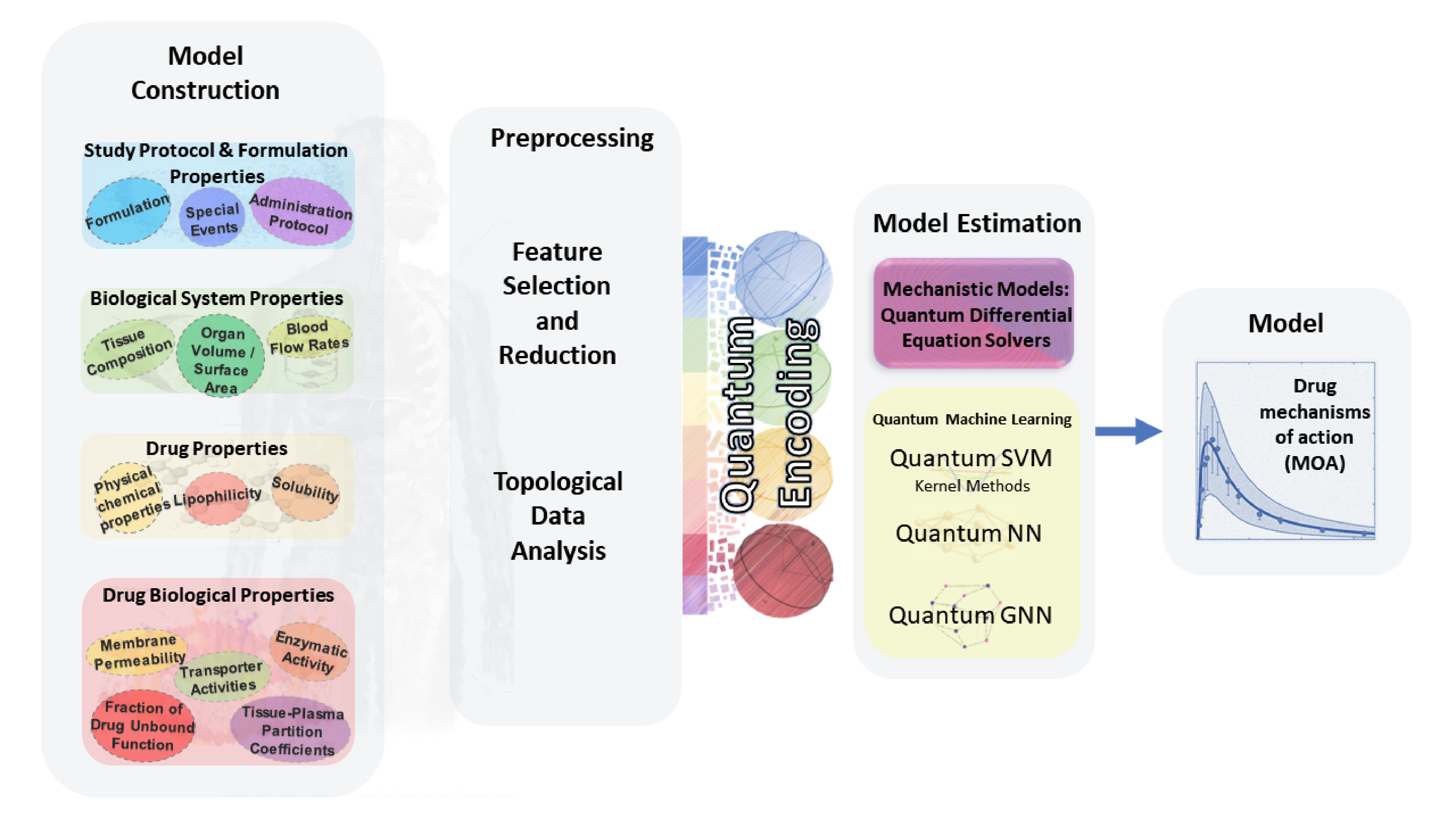}
    \caption{Schematic Representation of the PBPK Model Development and Estimation Workflow for Clinical Trials: The process begins with the construction of the model, incorporating drug properties and biological system attributes. It proceeds with data preprocessing, including feature selection, dimensionality reduction and topological data analysis. The final phase involves model estimation, leveraging quantum computational techniques such as quantum differential equation solvers, QSVMs, QNNs, and QGNNs, leading to the formulation of the model.}
    \label{trial-sim}
\end{figure}

PBPK models often require solving systems of ordinary differential equations (ODEs), a task that demands considerable computational resources, particularly for drug interaction simulations within intricate physiological contexts.  Quantum computing has been identified as a promising solution for accelerating both linear \cite{harrow2009quantum} and non-linear \cite{lloyd2020quantum, liu2021efficient, krovi2023improved} equations. Quantum differential equation solvers offer the potential to simulate PBPK models by solving differential equations that represent the non-linear and time-varying behaviour of drugs in the body. Furthermore, their ability to handle large-scale models, where the body is divided into several compartments, such as the liver, kidneys, lungs, and other organs appears very promising. Nonetheless, the deployment of such quantum algorithms is constrained by the current state of quantum hardware, which is yet to reach the necessary level of advancement. During this transitional phase, variational algorithms \cite{cerezo2021variational} for quantum machine learning and quantum kernel methods emerge as more viable options for the existing quantum hardware, albeit with caution \cite{cerezo2022challenges_QML,cerezo2023does, thanasilp2022exponential}. Besides the algorithms discussed in Section \ref{quantum}, research \cite{jaderberg2023let} has shown that integrating trainable parameters within data-encoding Hamiltonians could significantly improve the model's flexibility and forecasting precision. This strategy might allow for the fine-tuning of model frequencies for PBPK by enhancing the representation of complex drug distribution and metabolism processes. Additionally, the recent development of a spectral method which introduces an efficient technique for solving differential equations in high-dimensional spaces \cite{paine2023quantum, paine2023physics} by utilising a robust set of fitting polynomials, providing substantial expressiveness.

\subsection{Site Selection}

Selecting sites for a clinical trial can be formulated as an optimization problem. A straightforward approach is first determining which sites are feasible for the trial, followed by ranking the ``best'' sites according to its features and assigned weighted scores. The features are usually site specific historical performance data, and based on the final ranking of the sites, certain tiers are constructed to make a decision. These tiers can reflect the risks associated with the sites, namely the expense, enrollment rate and availability of expert PIs. A novel portfolio-based optimization approach is presented in \cite{site_portfolio_opt} as an improvement over the traditional approach. This more flexible model is designed to accommodate the study goals of the trial. Instead of ranking the sites and making a selection based on this ranking, portfolio-based optimization creates a portfolio of sites where the overall weight of the sites is optimal. This allows the model to consider ``what if" scenarios, trial specific constraints such as multiple sites in a country or preferred sites. A similar approach can be taken using quantum optimization algorithms. In particular, the problem formulation as a mixed-integer programming (MIP) can be mapped to a quantum setting. Interestingly, it is also possible to map the financial portfolio optimization quantum algorithm in \cite{rebentrost2018quantum} to clinical trial site selection setting. In this one-to-one mapping of the problem, we have the usual binary decision variables $x_i \in \{ 0,1\}$ to determine if the sites are selected in the portfolio or not. The quantum optimization can be set as:

\[
\min_{x\in \{0,1\}^n} qx^T \Sigma x - R^Tx \]
\[\text{subject to:} \,  A^Tx =S 
\]
where $R \in \mathbb{R}^n$ is the expected enrollment rate for trial sites, $\Sigma \in \mathbb{R}^{n \times n}$ is the covariance between trial sites, $q>0$ captures the risk that the trial design can specify which may be informed by trial specific constraints, $A \in \mathbb{R}^{n \times n}$ declares the trial constraints and $S$ denotes the number of sites to be selected in each portfolio. This optimization problem can be solved using either a Variational Quantum Eigensolver (VQE) or a Quantum Approximate Optimization Algorithm (QAOA) in near-term quantum computers. In principal, the ground state of the Hamiltonian returns the selection of a portfolio of sites. In the case of a high-dimensional parameter space where one considers all the sites in a country, or if the optimization problem exhibits non-convex behavior, we expect the quantum portfolio optimization method to perform well. For fault-tolerant quantum computers, it is shown in \cite{rebentrost2018quantum} that the proposed algorithm can solve the portfolio optimization in poly(log($N$)) run time compared to classical poly($N$), where $N$ is the total number of sites in our mapping. 

\begin{figure}[h!]
    \centering
    \includegraphics[width=17cm]{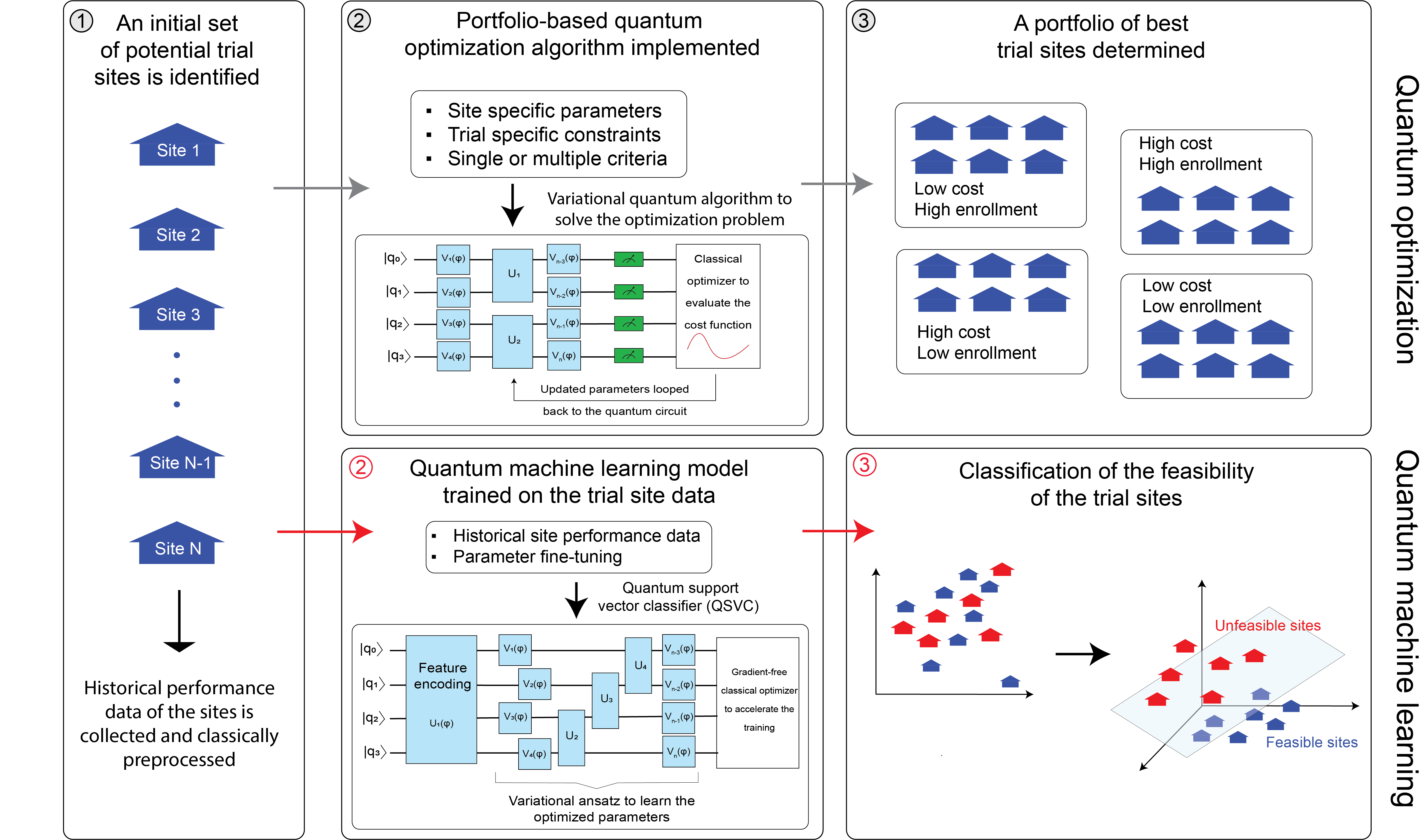}
    \caption{After the site-related data is collected and processed, we propose two approaches that utilize quantum computing. Portfolio-based optimization maps the problem with site-specific parameters and trial-specific constraints to select the best portfolio of sites. If the historical site performance data is limited, we propose in parallel, one can set a QSVC to classify which sites are feasible.}
    \label{site-selection}
\end{figure}

Another important bottleneck for site selection tasks is the availability of historical performance data from the potential trial sites. While some publicly available data can be used to identify sites, these are usually limited and can lack crucial information for optimal site selection. This can potentially create unwanted bias in the site selection process, impacting the cohort selection and the successful completion of the trial. As reported in \cite{theodorou2023framm}, missing modalities in the data can impact the site selection process. Quantum kernel methods \cite{schuld2019quantum_kernel} that map the feature space to a higher dimensional space to classify the data have been used in various QML applications. In particular, the work \cite{Kernel_method_EHR} demonstrates some significant results for QSVM for electronic health records (EHR) data and constructs a framework for empirical quantum advantage (EQA). The data they use is of similar nature to clinical trial site data, in the sense that samples are limited with high dimensional feature space. In this setting, we propose to use QSVM \cite{havlivcek2019supervised} for classifying the sites into feasible/not feasible classes. In particular, we expect quantum to provide utility where the sample-feature ratio is low with a high terrain ruggedness index as defined in \cite{Kernel_method_EHR}. This can potentially help overcome the bias towards selecting the same sites for running clinical trials, analyze the outcome of trials at new sites and provide access to more diverse cohorts.

\subsection{Cohort Identification}
Machine learning algorithms have shown promise for different aspects of cohort identification and selection, as well as aiding several enrichment strategies like prognostic and predictive enrichment to lower the failure rate and the required number of participants need to be recruited for clinical trials. However, a potential challenge of applying machine learning techniques to identify the cohorts based on the eligibility criteria is the diverse nature of the dataset. Classically several strategies have been developed known as fusion techniques to process diverse dataset like EHR, genomics, proteomics, and longitudinal data \cite{Kline2022}. Quantum feature map can potentially present a pathway to embed different biomedical data types more efficiently to train different quantum machine learning models ranging from supervised (e.g. QSVM \cite{Kernel_method_EHR, article_EHR, Ghoabdi2023.08.09.552597}), and unsupervised methods (e.g. quantum neural network\cite{mathur2022medical}). A potential development within the field of quantum-encoding that can impact cohort selection is the research of a universal feature map that can encode various data types ranging from images, EHR and gene expression to identify sub-population that has the highest positive response to certain clinical treatments. Such an advancement can provide unique benefits in predictive/prognosis enrichment strategy for high value  trials like oncology \cite{Van_Heertum2015-rv}, immunotherapy \cite{Limeta2023-jn} or cardiovascular diseases \cite{biom11111597}  since the usage of multimodal data (EHR, diagnostic imaging, multi-omics) can enhance the accuracy of ML models in predicting the efficacy of the treatments . 

Quantum feature map is a unique method to map classical data into the complex Hilbert space, which in theory can represent correlation between variables in complex dataset that is hard to simulate classically \cite{Huang2021, Havlíček2019}. Specifically, some feature maps involve constructing different forms of parameterized quantum circuits which encode the classical feature values. It has been shown that by encoding classical data in certain quantum circuit structure, this can induce correlation that does not exist in the dataset, but mimicking the behaviour of certain quantum states, thus potentially informing the performance of certain quantum machine learning algorithms in downstream applications like classification or regression \cite{kirk2023emergent}. 

\mycomment{Various quantum feature maps have been developed, among them are angle and amplitude encoding, Pauli Feature map, ZZFeature map, and IQP encodings \cite{rath2023quantum, Bremner_2016, Havlíček2019}.}

Another promising application of the quantum machine learning comes in the form of equivariant quantum neural network \cite{ragone2023representation, nguyen2022theory}. Equivariant quantum neural networks (EQNNs), also known as geometric quantum machine learning (GQML), is a class of quantum neural networks where the design of the quantum circuits contain the symmetries of the original dataset through the process of inductive bias \cite{Kbler2021TheIB}. Specifically, the quantum circuits are constructed based on the symmetry groups of the original datasets. EQNNs are designed to solve certain challenges of quantum neural networks such as barren plateau and excessive local minima, inhibiting their applications for different datasets \cite{Schatzki2024}. For biomedical dataset, it is expected that EQNNs can have important impact due to the various symmetries that intrinsically exist in different data modalities. For instance, medical images are known to contain symmetries like translation equivariance, rotation and reflection  \cite{10.1007/978-3-030-00934-2_24}, while sing-cell transcriptomics can contain unique local and global symmetrical structures \cite{Kobak2019}. As a result, EQNNs can be used to classify multi-model dataset to identify the right populations for clinical trials based on certain eligibility criteria. 

In addition to the usage QNNs for machine learning task like classification, a class of QNN known as quantum generative adversarial neural network (QGAN) has gained significant interest due to their unique advantages. Quantum GAN has been proposed as an alternative to overcome many disadvantages of classical GAN. For instance, it has been proved that QGAN will require less training data to produce robust model \cite{Caro2022}. There are many different implementation of the QGAN such as hybrid model with quantum generator and classical discriminator, as well as a full quantum version with QNNs acting as generator and discriminator. In the fully quantum settings, the quantum GAN exhibits a unique advantage with the optimization between the generator and discriminator is a linear programming optimization, with a convex state-space as described by the Kakutani fixed-point theorem \cite{PhysRevLett.121.040502}. In practice, full quantum GAN has been used to produce simple synthetic images based on the MNIST dataset \cite{Stein_2021}, while hybrid versions with quantum generator and classical discriminator have been used to synthesize random distributions \cite{Zoufal2019}. As a result, by using either a hybrid QGAN or a full quantum version, these setups can generate complex distribution mimicking the distribution from real patient biomedical data with much lower training data, acting as synthetic control arm for clinical trial. The usage of QGAN can potentially benefit rare disease trial since the training data for the generative model tends to be small due to complex nature of the disease. 

\begin{figure}[h!]
    \centering
    \includegraphics[width=17cm]{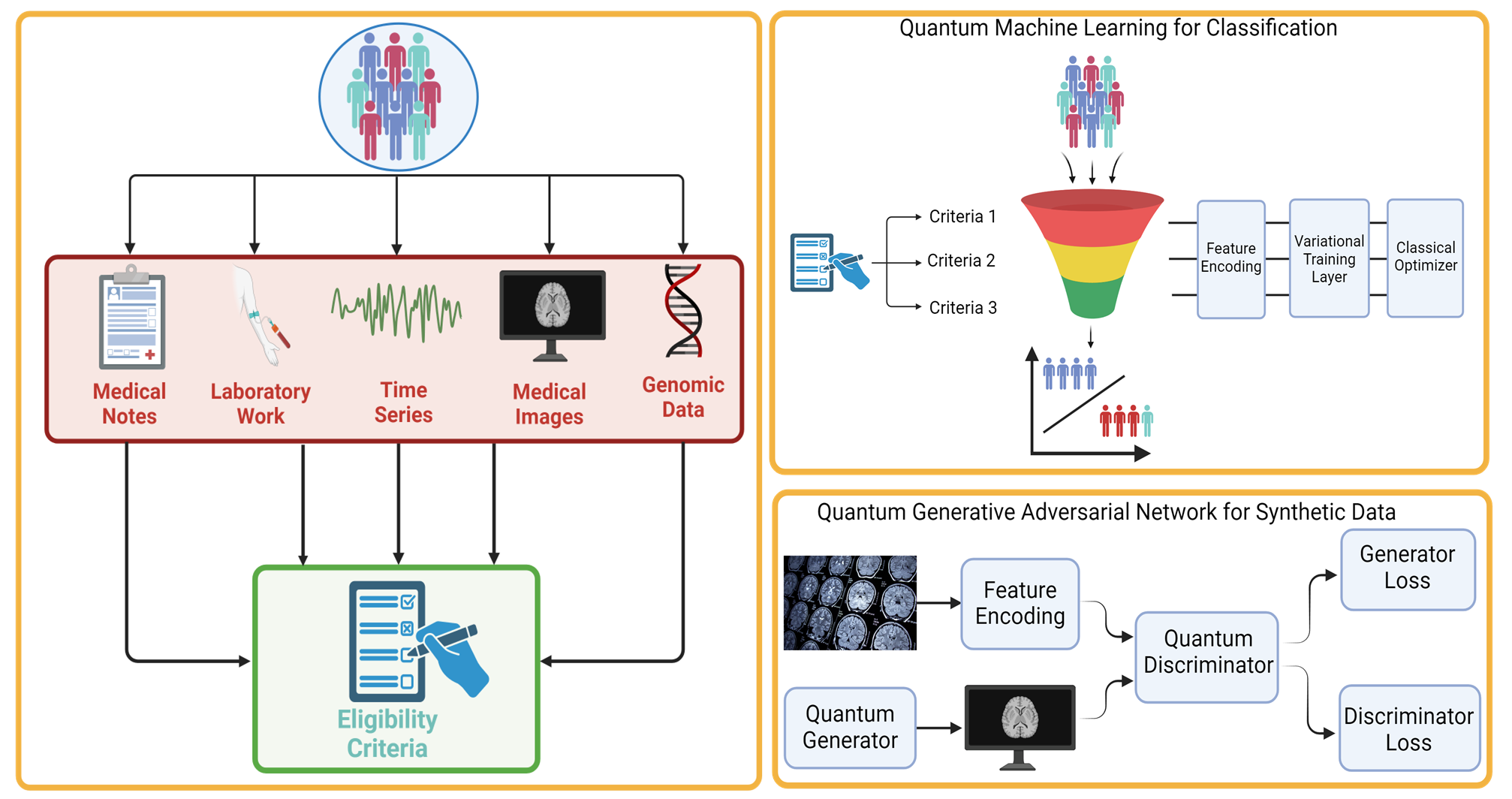}
    \caption{Overview of the cohort identification process with multi-modal data, and applications of quantum machine learning for classification and generative model in the cohort identification process.}
    \label{cohort-identification}
\end{figure}

\section{Prospect of quantum computing in clinical trial design and optimization}

Our work proposes various quantum algorithms to improve the outcome of clinical trials, and discusses in depth how practitioners of the field can utilize these new computational tools. By simulating the drug properties more accurately, we can determine the trial protocol that can ensure a successful outcome. Furthermore, optimizing the procedures of site selection and cohort identification based on the protocol design, we can reduce the cost of the clinical trials and prevent adverse events. As quantum computing technology matures, we see utility-scale quantum experiments displaying real promise to advance our computational capabilities \cite{kim2023evidence}. In parallel with that, more efficient quantum error correction methods \cite{bravyi2024high} bring the arrival of fault-tolerant quantum computers closer than expected. Without a doubt, quantum computing will play an important role to advance science in the next decade, and this work explores the potential of quantum computing to enhance clinical trial outcomes.   

% Mention quantum sensing even if not with a lot of detail?

\bibliographystyle{unsrt}
\bibliography{bibliography}

\begin{thebibliography}{100}

\bibitem{dirac1981principles}
Paul Adrien~Maurice Dirac.
\newblock The principles of quantum mechanics, 1981.

\bibitem{heisenberg1973development}
Werner Heisenberg.
\newblock Development of concepts in the history of quantum theory.
\newblock In {\em The physicist’s conception of nature}, pages 264--275.
  Springer, 1973.

\bibitem{shor1994algorithms}
Peter~W Shor.
\newblock Algorithms for quantum computation: discrete logarithms and
  factoring.
\newblock In {\em Proceedings 35th annual symposium on foundations of computer
  science}, pages 124--134. Ieee, 1994.

\bibitem{grover1996fast}
Lov~K Grover.
\newblock A fast quantum mechanical algorithm for database search.
\newblock In {\em Proceedings of the twenty-eighth annual ACM symposium on
  Theory of computing}, pages 212--219, 1996.

\bibitem{harrow2009quantum}
Aram~W Harrow, Avinatan Hassidim, and Seth Lloyd.
\newblock Quantum algorithm for linear systems of equations.
\newblock {\em Physical review letters}, 103(15):150502, 2009.

\bibitem{aharonov2006polynomial}
Dorit Aharonov, Vaughan Jones, and Zeph Landau.
\newblock A polynomial quantum algorithm for approximating the jones
  polynomial.
\newblock In {\em Proceedings of the thirty-eighth annual ACM symposium on
  Theory of computing}, pages 427--436, 2006.

\bibitem{lloyd2016quantum}
Seth Lloyd, Silvano Garnerone, and Paolo Zanardi.
\newblock Quantum algorithms for topological and geometric analysis of data.
\newblock {\em Nature communications}, 7(1):10138, 2016.

\bibitem{lloyd1996universal}
Seth Lloyd.
\newblock Universal quantum simulators.
\newblock {\em Science}, 273(5278):1073--1078, 1996.

\bibitem{clinicalfailurerate}
Sun D, Gao W, Hu~H, and Zhou S.
\newblock Why 90% of clinical drug development fails and how to improve it?
\newblock {\em Acta Pharm Sin B.}, 2022.

\bibitem{Harrer_2019_AI_CTs}
Stefan Harrer, Pratik Shah, Bhavna Antony, and Jianying Hu.
\newblock Artificial intelligence for clinical trial design.
\newblock {\em Trends in pharmacological sciences}, 40(8):577--591, 2019.

\bibitem{site_Njoku_2023}
Kingsley Njoku, Tesfaye Yadete, Joao Bettencourt-Silva, Matthew~T. Liu, Vibha
  Anand, Alberto Purpura, Uri Kartoun, Eileen Koski, Natasha Mulligan, Brian
  Claggett, Thaddeus Stappenbeck, and Julia~J. Liu.
\newblock Clinical trial failures in inflammatory bowel disease: An artificial
  intelligence-assisted review.
\newblock {\em Official journal of the American College of Gastroenterology
  {\textbar} {ACG}}, 118(10), 2023.

\bibitem{site_Lamberti_2021}
Mary~Jo Lamberti, Zachary Smith, Rhonda Henry, Deborah Howe, Melanie Goodwin,
  Amanda Williams, and Kenneth Getz.
\newblock Benchmarking patient recruitment and retention practices.
\newblock {\em Therapeutic Innovation \& Regulatory Science}, 55(1):19--32,
  2021.

\bibitem{FDA_2019_Trial_Enrichment}
Food and Drug Administration.
\newblock Enrichment strategies for clinical trials to support approval of
  human drugs and biological products.
\newblock Technical report, Center for Drug Evaluation and Research, 2019.

\bibitem{site_Fogel_2018}
David~B. Fogel.
\newblock Factors associated with clinical trials that fail and opportunities
  for improving the likelihood of success: A review.
\newblock {\em Contemp Clin Trials Commun}, 11:156--164, 2018.
\newblock Place: Netherlands.

\bibitem{dimasi2016innovation}
Joseph~A DiMasi, Henry~G Grabowski, and Ronald~W Hansen.
\newblock Innovation in the pharmaceutical industry: new estimates of r\&d
  costs.
\newblock {\em Journal of health economics}, 47:20--33, 2016.

\bibitem{wong2019estimation}
Chi~Heem Wong, Kien~Wei Siah, and Andrew~W Lo.
\newblock Estimation of clinical trial success rates and related parameters.
\newblock {\em Biostatistics}, 20(2):273--286, 2019.

\bibitem{wouters2020estimated}
Olivier~J Wouters, Martin McKee, and Jeroen Luyten.
\newblock Estimated research and development investment needed to bring a new
  medicine to market, 2009-2018.
\newblock {\em Jama}, 323(9):844--853, 2020.

\bibitem{otava2015identification}
Martin Otava, Ziv Shkedy, Willem Talloen, Geert~R Verheyen, and Adetayo Kasim.
\newblock Identification of in vitro and in vivo disconnects using
  transcriptomic data.
\newblock {\em BMC genomics}, 16:1--10, 2015.

\bibitem{gonzalez2015vitro}
Ignacio Gonz{\'a}lez-Garc{\'\i}a, Victor Mangas-Sanju{\'a}n, Matilde
  Merino-Sanju{\'a}n, and Marival Bermejo.
\newblock In vitro--in vivo correlations: general concepts, methodologies and
  regulatory applications.
\newblock {\em Drug development and industrial pharmacy}, 41(12):1935--1947,
  2015.

\bibitem{kim2020human}
Jihoon Kim, Bon-Kyoung Koo, and Juergen~A Knoblich.
\newblock Human organoids: model systems for human biology and medicine.
\newblock {\em Nature Reviews Molecular Cell Biology}, 21(10):571--584, 2020.

\bibitem{hamburg2010path}
Margaret~A Hamburg and Francis~S Collins.
\newblock The path to personalized medicine.
\newblock {\em New England Journal of Medicine}, 363(4):301--304, 2010.

\bibitem{site_Bieske_2023}
Linn Bieske, Maximillian Zinner, Florian Dahlhausen, and Hubert Truebel.
\newblock Critical path activities in clinical trial setup and conduct: How to
  avoid bottlenecks and accelerate clinical trials.
\newblock {\em Drug Discovery Today}, 28(10):103733, 2023.

\bibitem{Strict_criteria_oncology}
Nessa Stefaniak, Jennifer Walker, Monica~L. Murphy, Mishellene McKinney,
  Lu~Liu, and Stephen~B. Edge.
\newblock Do eligibility criteria restrict access to clinical trials?
\newblock {\em Journal of Clinical Oncology}, 38(29\_suppl):94--94, 2020.

\bibitem{10.1001/jama.297.11.1233}
Harriette G.~C. Van~Spall, Andrew Toren, Alex Kiss, and Robert~A. Fowler.
\newblock {Eligibility Criteria of Randomized Controlled Trials Published in
  High-Impact General Medical JournalsA Systematic Sampling Review}.
\newblock {\em JAMA}, 297(11):1233--1240, 03 2007.

\bibitem{Generalization_criteria}
Edward~S. Kim, Suanna~S. Bruinooge, Samantha Roberts, Gwynn Ison, Nancy~U. Lin,
  Lia Gore, Thomas~S. Uldrick, Stuart~M. Lichtman, Nancy Roach, Julia~A.
  Beaver, Rajeshwari Sridhara, Paul~J. Hesketh, Andrea~M. Denicoff, Elizabeth
  Garrett-Mayer, Eric Rubin, Pratik Multani, Tatiana~M. Prowell, Caroline
  Schenkel, Marina Kozak, Jeff Allen, Ellen Sigal, and Richard~L. Schilsky.
\newblock Broadening eligibility criteria to make clinical trials more
  representative: American society of clinical oncology and friends of cancer
  research joint research statement.
\newblock {\em Journal of Clinical Oncology}, 35(33):3737--3744, 2017.
\newblock PMID: 28968170.

\bibitem{10.1001/jama.2019.17016}
Jennifer Abbasi.
\newblock {Older Patients (Still) Left Out of Cancer Clinical Trials}.
\newblock {\em JAMA}, 322(18):1751--1753, 11 2019.

\bibitem{10.6004/jnccn.2017.7050}
Juno Obedin-Maliver.
\newblock Time to change: Supporting sexual and gender minority people—an
  underserved, understudied cancer risk population.
\newblock {\em Journal of the National Comprehensive Cancer Network J Natl
  Compr Canc Netw}, 15(11):1305 -- 1308, 2017.

\bibitem{doi:10.1177/107327481602300404}
Lauren~M. Hamel, Louis~A. Penner, Terrance~L. Albrecht, Elisabeth Heath,
  Clement~K. Gwede, and Susan Eggly.
\newblock Barriers to clinical trial enrollment in racial and ethnic minority
  patients with cancer.
\newblock {\em Cancer Control}, 23(4):327--337, 2016.
\newblock PMID: 27842322.

\bibitem{FDA_guideline}
Food and Drug Aministration.
\newblock Enhancing the diversity of clinical trial populations — eligibility
  criteria, enrollment practices, and trial designs guidance for industry.
\newblock Technical report, 2020.

\bibitem{doi:10.1200/JCO.2015.62.1854}
Edward~S. Kim, David Bernstein, Susan~G. Hilsenbeck, Christine~H. Chung,
  Adam~P. Dicker, Jennifer~L. Ersek, Steven Stein, Fadlo~R. Khuri, Earle
  Burgess, Kelly Hunt, Percy Ivy, Suanna~S. Bruinooge, Neal Meropol, and
  Richard~L. Schilsky.
\newblock Modernizing eligibility criteria for molecularly driven trials.
\newblock {\em Journal of Clinical Oncology}, 33(25):2815--2820, 2015.
\newblock PMID: 26195710.

\bibitem{yekeduz2021assessing}
Emre Yeked{\"u}z, Dario Trapani, Wenxin Xu, Elisabeth~GE de~Vries, Chris
  Labaki, Bishal Gyawali, Shuchi Gulati, Chadi Nabhan, G{\"u}ng{\"o}r Utkan,
  Giuseppe Curigliano, et~al.
\newblock Assessing population diversity in phase iii trials of cancer drugs
  supporting food and drug administration approval in solid tumors.
\newblock {\em International journal of cancer}, 149(7):1455--1462, 2021.

\bibitem{khoury2022health}
Muin~J Khoury, Scott Bowen, W~David Dotson, Emily Drzymalla, Ridgely~F Green,
  Robert Goldstein, Katherine Kolor, Leandris~C Liburd, Laurence~S Sperling,
  and Rebecca Bunnell.
\newblock Health equity in the implementation of genomics and precision
  medicine: A public health imperative.
\newblock {\em Genetics in Medicine}, 24(8):1630--1639, 2022.

\bibitem{zaaijer2021ancestry}
Sophie Zaaijer and Amanda Capes-Davis.
\newblock Ancestry matters: Building inclusivity into preclinical study design.
\newblock {\em Cell}, 184(10):2525--2531, 2021.

\bibitem{RODRIGUEZTORRES2021100829}
Edgardo Rodríguez-Torres, Margarita~M. González-Pérez, and Clemente
  Díaz-Pérez.
\newblock Barriers and facilitators to the participation of subjects in
  clinical trials: An overview of reviews.
\newblock {\em Contemporary Clinical Trials Communications}, 23:100829, 2021.

\bibitem{Stanski2020}
Natalja~L. Stanski and Hector~R. Wong.
\newblock Prognostic and predictive enrichment in sepsis.
\newblock {\em Nature Reviews Nephrology}, 16(1):20--31, Jan 2020.

\bibitem{Wong2016-ao}
Hector~R Wong, Sarah~J Atkinson, Natalie~Z Cvijanovich, Nick Anas, Geoffrey~L
  Allen, Neal~J Thomas, Michael~T Bigham, Scott~L Weiss, Julie~C Fitzgerald,
  Paul~A Checchia, Keith Meyer, Michael Quasney, Mark Hall, Rainer Gedeit,
  Robert~J Freishtat, Jeffrey Nowak, Shekhar~S Raj, Shira Gertz, and
  Christopher~J Lindsell.
\newblock Combining prognostic and predictive enrichment strategies to identify
  children with septic shock responsive to corticosteroids.
\newblock {\em Crit. Care Med.}, 44(10):e1000--3, October 2016.

\bibitem{10.1117/12.2178956}
Stefan Harrer.
\newblock {Measuring life: sensors and analytics for precision medicine}.
\newblock In Sander van~den Driesche, editor, {\em Bio-MEMS and Medical
  Microdevices II}, volume 9518, page 951802. International Society for Optics
  and Photonics, SPIE, 2015.

\bibitem{10.1136/amiajnl-2013-001935}
Chaitanya Shivade, Preethi Raghavan, Eric Fosler-Lussier, Peter~J Embi, Noemie
  Elhadad, Stephen~B Johnson, and Albert~M Lai.
\newblock {A review of approaches to identifying patient phenotype cohorts
  using electronic health records}.
\newblock {\em Journal of the American Medical Informatics Association},
  21(2):221--230, 11 2013.

\bibitem{10.1093/biostatistics/kxn010}
Sara Geneletti, Sylvia Richardson, and Nicky Best.
\newblock {Adjusting for selection bias in retrospective, case–control
  studies}.
\newblock {\em Biostatistics}, 10(1):17--31, 05 2008.

\bibitem{spasic2019cohort}
Irena Spasic, Dominik Krzeminski, Padraig Corcoran, Alexander Balinsky, et~al.
\newblock Cohort selection for clinical trials from longitudinal patient
  records: text mining approach.
\newblock {\em JMIR medical informatics}, 7(4):e15980, 2019.

\bibitem{Wells2013StrategiesFH}
Brian~J. Wells, Kevin Chagin, Amy~S. Nowacki, and Michael~W. Kattan.
\newblock Strategies for handling missing data in electronic health record
  derived data.
\newblock {\em eGEMs}, 1, 2013.

\bibitem{Rockowitz2020}
Shira Rockowitz, Nicholas LeCompte, Mary Carmack, Andrew Quitadamo, Lily Wang,
  Meredith Park, Devon Knight, Emma Sexton, Lacey Smith, Beth Sheidley, Michael
  Field, Ingrid~A. Holm, Catherine~A. Brownstein, Pankaj~B. Agrawal, Susan
  Kornetsky, Annapurna Poduri, Scott~B. Snapper, Alan~H. Beggs, Timothy~W. Yu,
  David~A. Williams, and Piotr Sliz.
\newblock Children's rare disease cohorts: an integrative research and clinical
  genomics initiative.
\newblock {\em npj Genomic Medicine}, 5(1):29, Jul 2020.

\bibitem{Kline2022}
Adrienne Kline, Hanyin Wang, Yikuan Li, Saya Dennis, Meghan Hutch, Zhenxing Xu,
  Fei Wang, Feixiong Cheng, and Yuan Luo.
\newblock Multimodal machine learning in precision health: A scoping review.
\newblock {\em npj Digital Medicine}, 5(1):171, Nov 2022.

\bibitem{Hee2017}
Siew~Wan Hee, Adrian Willis, Catrin Tudur~Smith, Simon Day, Frank Miller, Jason
  Madan, Martin Posch, Sarah Zohar, and Nigel Stallard.
\newblock Does the low prevalence affect the sample size of interventional
  clinical trials of rare diseases? an analysis of data from the aggregate
  analysis of clinicaltrials.gov.
\newblock {\em Orphanet Journal of Rare Diseases}, 12(1):44, Mar 2017.

\bibitem{synthetic_arms_thorlund2020}
Kristian Thorlund, Louis Dron, Jay~JH Park, and Edward~J Mills.
\newblock Synthetic and external controls in clinical trials--a primer for
  researchers.
\newblock {\em Clinical epidemiology}, pages 457--467, 2020.

\bibitem{ML_PAI}
Suzanne~C. van Bronswijk, Sanne J.~E. Bruijniks, Lorenzo Lorenzo-Luaces,
  Robert~J. Derubeis, Lotte H. J.~M. Lemmens, Frenk P. M.~L. Peeters, and
  Marcus. J.~H. Huibers.
\newblock Cross-trial prediction in psychotherapy: External validation of the
  personalized advantage index using machine learning in two dutch randomized
  trials comparing cbt versus ipt for depression.
\newblock {\em Psychotherapy Research}, 31(1):78--91, 2021.
\newblock PMID: 32964809.

\bibitem{Mellem2021}
Monika~S. Mellem, Matt Kollada, Jane Tiller, and Thomas Lauritzen.
\newblock Explainable ai enables clinical trial patient selection to
  retrospectively improve treatment effects in schizophrenia.
\newblock {\em BMC Medical Informatics and Decision Making}, 21(1):162, May
  2021.

\bibitem{gutierrez2021methods}
Jos{\'e}~Ram{\'o}n Guti{\'e}rrez-Casares, Javier Quintero, Guillem Jorba,
  Valentin Junet, Vicente Mart{\'\i}nez, Tamara Pozo-Rubio, Baldomero Oliva,
  Xavier Daura, Jos{\'e}~Manuel Mas, and Carmen Montoto.
\newblock Methods to develop an in silico clinical trial: computational
  head-to-head comparison of lisdexamfetamine and methylphenidate.
\newblock {\em Frontiers in Psychiatry}, 12:741170, 2021.

\bibitem{inan2020digitizing}
OT~Inan, P~Tenaerts, SA~Prindiville, HR~Reynolds, DS~Dizon, K~Cooper-Arnold,
  M~Turakhia, MJ~Pletcher, KL~Preston, HM~Krumholz, et~al.
\newblock Digitizing clinical trials.
\newblock {\em NPJ digital medicine}, 3(1):101, 2020.

\bibitem{hekler2016advancing}
Eric~B Hekler, Susan Michie, Misha Pavel, Daniel~E Rivera, Linda~M Collins,
  Holly~B Jimison, Claire Garnett, Skye Parral, and Donna Spruijt-Metz.
\newblock Advancing models and theories for digital behavior change
  interventions.
\newblock {\em American journal of preventive medicine}, 51(5):825--832, 2016.

\bibitem{liao2016sample}
Peng Liao, Predrag Klasnja, Ambuj Tewari, and Susan~A Murphy.
\newblock Sample size calculations for micro-randomized trials in mhealth.
\newblock {\em Statistics in medicine}, 35(12):1944--1971, 2016.

\bibitem{kuepfer2016applied}
L~Kuepfer, C~Niederalt, T~Wendl, J-F Schlender, S~Willmann, J~Lippert, M~Block,
  T~Eissing, and D~Teutonico.
\newblock Applied concepts in pbpk modeling: how to build a pbpk/pd model.
\newblock {\em CPT: pharmacometrics \& systems pharmacology}, 5(10):516--531,
  2016.

\bibitem{hartmanshenn2016physiologically}
Clara Hartmanshenn, Megerle Scherholz, and Ioannis~P Androulakis.
\newblock Physiologically-based pharmacokinetic models: approaches for enabling
  personalized medicine.
\newblock {\em Journal of pharmacokinetics and pharmacodynamics}, 43:481--504,
  2016.

\bibitem{rowland2011physiologically}
Malcolm Rowland, Carl Peck, and Geoffrey Tucker.
\newblock Physiologically-based pharmacokinetics in drug development and
  regulatory science.
\newblock {\em Annual review of pharmacology and toxicology}, 51:45--73, 2011.

\bibitem{chetty2014applications}
Manoranjenni Chetty, Rachel~H Rose, Khaled Abduljalil, Nikunjkumar Patel,
  Gaohua Lu, Theresa Cain, Masoud Jamei, and Amin Rostami-Hodjegan.
\newblock Applications of linking pbpk and pd models to predict the impact of
  genotypic variability, formulation differences, differences in target binding
  capacity and target site drug concentrations on drug responses and
  variability.
\newblock {\em Frontiers in pharmacology}, 5:258, 2014.

\bibitem{jamei2009population}
Masoud Jamei, David Turner, Jiansong Yang, Sibylle Neuhoff, Sebastian Polak,
  Amin Rostami-Hodjegan, and Geoffrey Tucker.
\newblock Population-based mechanistic prediction of oral drug absorption.
\newblock {\em The AAPS journal}, 11:225--237, 2009.

\bibitem{huang2013utility}
Shiew-Mei Huang, Darrell~R Abernethy, Yaning Wang, Ping Zhao, and Issam Zineh.
\newblock The utility of modeling and simulation in drug development and
  regulatory review.
\newblock {\em Journal of pharmaceutical sciences}, 102(9):2912--2923, 2013.

\bibitem{kang2009overview}
Ju-Seop Kang and Min-Ho Lee.
\newblock Overview of therapeutic drug monitoring.
\newblock {\em The Korean journal of internal medicine}, 24(1):1, 2009.

\bibitem{strougo2011semiphysiological}
Ashley Strougo, Ashraf Yassen, Walter Krauwinkel, Meindert Danhof, and Jan
  Freijer.
\newblock A semiphysiological population model for prediction of the
  pharmacokinetics of drugs under liver and renal disease conditions.
\newblock {\em Drug metabolism and disposition}, 39(7):1278--1287, 2011.

\bibitem{gehring2013factors}
Marta Gehring, Rod~S Taylor, Marie Mellody, Brigitte Casteels, Angela Piazzi,
  Gianfranco Gensini, and Giuseppe Ambrosio.
\newblock Factors influencing clinical trial site selection in europe: the
  survey of attitudes towards trial sites in europe (the sat-eu study).
\newblock {\em BMJ open}, 3(11), 2013.

\bibitem{shao2020review}
Meng Shao, Zhixin Han, Jinwei Sun, Chengsi Xiao, Shulei Zhang, and Yuanxu Zhao.
\newblock A review of multi-criteria decision making applications for renewable
  energy site selection.
\newblock {\em Renewable Energy}, 157:377--403, 2020.

\bibitem{csener2006landfill}
Ba{\c{s}}ak {\c{S}}ener, M~L{\"u}tfi S{\"u}zen, and Vedat Doyuran.
\newblock Landfill site selection by using geographic information systems.
\newblock {\em Environmental geology}, 49:376--388, 2006.

\bibitem{liang1991fuzzy}
Gin-Shuh Liang and Mao-Jiun~J Wang.
\newblock A fuzzy multi-criteria decision-making method for facility site
  selection.
\newblock {\em The International Journal of Production Research},
  29(11):2313--2330, 1991.

\bibitem{site_portfolio_opt}
MD~Paluy, Vadim and PhD Shnaydman, Vladimir.
\newblock Portfolio approach to optimize site selection.
\newblock {\em Applied Clinical Trials}, 2019.

\bibitem{theodorou2023framm}
Brandon Theodorou, Lucas Glass, Cao Xiao, and Jimeng Sun.
\newblock Framm: Fair ranking with missing modalities for clinical trial site
  selection.
\newblock {\em arXiv preprint arXiv:2305.19407}, 2023.

\bibitem{jorion1992portfolio}
Philippe Jorion.
\newblock Portfolio optimization in practice.
\newblock {\em Financial analysts journal}, 48(1):68--74, 1992.

\bibitem{Hurtado-Chonge014796}
Anah{\'\i} Hurtado-Chong, Alexander Joeris, Denise Hess, and Michael Blauth.
\newblock Improving site selection in clinical studies: a standardised,
  objective, multistep method and first experience results.
\newblock {\em BMJ Open}, 7(7), 2017.

\bibitem{biswal2019doctor2vec}
Siddharth Biswal, Cao Xiao, Lucas~M. Glass, Elizabeth Milkovits, and Jimeng
  Sun.
\newblock Doctor2vec: Dynamic doctor representation learning for clinical trial
  recruitment.
\newblock {\em arXiv preprint arXiv:1911.10395}, 2019.

\bibitem{10.1093/jamia/ocz064}
Jelena Gligorijevic, Djordje Gligorijevic, Martin Pavlovski, Elizabeth
  Milkovits, Lucas Glass, Kevin Grier, Praveen Vankireddy, and Zoran Obradovic.
\newblock {Optimizing clinical trials recruitment via deep learning}.
\newblock {\em Journal of the American Medical Informatics Association},
  26(11):1195--1202, 06 2019.

\bibitem{srinivasa2022clinical}
Rakshith~S Srinivasa, Cheng Qian, Brandon Theodorou, Jeffrey Spaeder, Cao Xiao,
  Lucas Glass, and Jimeng Sun.
\newblock Clinical trial site matching with improved diversity using fair
  policy learning.
\newblock {\em arXiv preprint arXiv:2204.06501}, 2022.

\bibitem{wang2023pytrial}
Zifeng Wang, Brandon Theodorou, Tianfan Fu, Cao Xiao, and Jimeng Sun.
\newblock Pytrial: Machine learning software and benchmark for clinical trial
  applications.
\newblock {\em arXiv preprint arXiv:2306.04018}, 2023.

\bibitem{site_CTKG}
Chen Z, Peng B, Ioannidis VN, Li~M, Karypis G, and Ning X.
\newblock A knowledge graph of clinical trials.
\newblock {\em Sci Rep}, 2022.

\bibitem{cui2023survey}
Hejie Cui, Jiaying Lu, Shiyu Wang, Ran Xu, Wenjing Ma, Shaojun Yu, Yue Yu, Xuan
  Kan, Chen Ling, Joyce Ho, et~al.
\newblock A survey on knowledge graphs for healthcare: Resources, applications,
  and promises.
\newblock {\em arXiv preprint arXiv:2306.04802}, 2023.

\bibitem{segura2019cohort}
Isabel Segura-Bedmar and Pablo Raez.
\newblock Cohort selection for clinical trials using deep learning models.
\newblock {\em Journal of the American Medical Informatics Association},
  26(11):1181--1188, 2019.

\bibitem{chen2023machine}
Tony Lin-Wei Chen, Anirudh Buddhiraju, Timothy~G Costales, Murad~Abdullah
  Subih, Henry~Hojoon Seo, and Young-Min Kwon.
\newblock Machine learning models based on a national-scale cohort identify
  patients at high risk of prolonged length of stay following primary total hip
  arthroplasty.
\newblock {\em The Journal of Arthroplasty}, 2023.

\bibitem{krishnasamy2014hybrid}
Ganesh Krishnasamy, Anand~J Kulkarni, and Raveendran Paramesran.
\newblock A hybrid approach for data clustering based on modified cohort
  intelligence and k-means.
\newblock {\em Expert Systems with Applications}, 41(13):6009--6016, 2014.

\bibitem{li2020behrt}
Yikuan Li, Shishir Rao, Jos{\'e} Roberto~Ayala Solares, Abdelaali Hassaine,
  Rema Ramakrishnan, Dexter Canoy, Yajie Zhu, Kazem Rahimi, and Gholamreza
  Salimi-Khorshidi.
\newblock Behrt: transformer for electronic health records.
\newblock {\em Scientific reports}, 10(1):7155, 2020.

\bibitem{goodfellow2020generative}
Ian Goodfellow, Jean Pouget-Abadie, Mehdi Mirza, Bing Xu, David Warde-Farley,
  Sherjil Ozair, Aaron Courville, and Yoshua Bengio.
\newblock Generative adversarial networks.
\newblock {\em Communications of the ACM}, 63(11):139--144, 2020.

\bibitem{KRENMAYR2022101118}
Lucas Krenmayr, Roland Frank, Christina Drobig, Michael Braungart, Jan Seidel,
  Daniel Schaudt, Reinhold {von Schwerin}, and Kathrin Stucke-Straub.
\newblock Ganeraid: Realistic synthetic patient data for clinical trials.
\newblock {\em Informatics in Medicine Unlocked}, 35:101118, 2022.

\bibitem{platt2024creating}
Daniel~Enoch Platt, Aritra Bose, Kahn Rhrissorrakrai, Aldo~Guzman Saenz, Niina
  Haiminen, and Laxmi Parida.
\newblock Creating synthetic patient data using a generative adversarial
  network having a multivariate gaussian generative model, March~14 2024.
\newblock US Patent App. 17/930,477.

\bibitem{10.1145/3422622}
Ian Goodfellow, Jean Pouget-Abadie, Mehdi Mirza, Bing Xu, David Warde-Farley,
  Sherjil Ozair, Aaron Courville, and Yoshua Bengio.
\newblock Generative adversarial networks.
\newblock {\em Commun. ACM}, 63(11):139–144, oct 2020.

\bibitem{Soenksen2022}
Luis~R. Soenksen, Yu~Ma, Cynthia Zeng, Leonard Boussioux, Kimberly
  Villalobos~Carballo, Liangyuan Na, Holly~M. Wiberg, Michael~L. Li, Ignacio
  Fuentes, and Dimitris Bertsimas.
\newblock Integrated multimodal artificial intelligence framework for
  healthcare applications.
\newblock {\em npj Digital Medicine}, 5(1):149, Sep 2022.

\bibitem{mohsen2022artificial}
Farida Mohsen, Hazrat Ali, Nady El~Hajj, and Zubair Shah.
\newblock Artificial intelligence-based methods for fusion of electronic health
  records and imaging data.
\newblock {\em Scientific Reports}, 12(1):17981, 2022.

\bibitem{fountzilas2022clinical}
Elena Fountzilas, Apostolia~M Tsimberidou, Henry~Hiep Vo, and Razelle Kurzrock.
\newblock Clinical trial design in the era of precision medicine.
\newblock {\em Genome medicine}, 14(1):101, 2022.

\bibitem{hu2019biomarker}
Chen Hu and James~J Dignam.
\newblock Biomarker-driven oncology clinical trials: key design elements,
  types, features, and practical considerations.
\newblock {\em JCO Precision Oncology}, 1:1--12, 2019.

\bibitem{Berchialla2022}
Paola Berchialla, Corrado Lanera, Veronica Sciannameo, Dario Gregori, and
  Ileana Baldi.
\newblock Prediction of treatment outcome in clinical trials under a
  personalized medicine perspective.
\newblock {\em Scientific Reports}, 12(1):4115, Mar 2022.

\bibitem{food2019drug}
US~Food.
\newblock Drug administration. enrichment strategies for clinical trials to
  support determination of effectiveness of human drugs and biological products
  guidance for industry.
\newblock {\em Administration FD, editor. Silver Spring: Center for Drug
  Evaluation and Research}, pages 1--41, 2019.

\bibitem{fahed2022potential}
Akl~C Fahed, Anthony~A Philippakis, and Amit~V Khera.
\newblock The potential of polygenic scores to improve cost and efficiency of
  clinical trials.
\newblock {\em Nature communications}, 13(1):2922, 2022.

\bibitem{natarajan2017polygenic}
Pradeep Natarajan, Robin Young, Nathan~O Stitziel, Sandosh Padmanabhan, Usman
  Baber, Roxana Mehran, Samantha Sartori, Valentin Fuster, Dermot~F Reilly,
  Adam Butterworth, et~al.
\newblock Polygenic risk score identifies subgroup with higher burden of
  atherosclerosis and greater relative benefit from statin therapy in the
  primary prevention setting.
\newblock {\em Circulation}, 135(22):2091--2101, 2017.

\bibitem{mega2015genetic}
Jessica~L Mega, Nathan~O Stitziel, J~Gustav Smith, Daniel~I Chasman, Mark~J
  Caulfield, James~J Devlin, Francesco Nordio, Craig~L Hyde, Christopher~P
  Cannon, Frank~M Sacks, et~al.
\newblock Genetic risk, coronary heart disease events, and the clinical benefit
  of statin therapy: an analysis of primary and secondary prevention trials.
\newblock {\em The Lancet}, 385(9984):2264--2271, 2015.

\bibitem{bose2023role}
Aritra Bose, Daniel~E Platt, Uri Kartoun, Kenney Ng, and Laxmi Parida.
\newblock Role of genetics in capturing racial disparities in cardiovascular
  disease.
\newblock {\em medRxiv}, pages 2023--02, 2023.

\bibitem{martin2019clinical}
Alicia~R Martin, Masahiro Kanai, Yoichiro Kamatani, Yukinori Okada, Benjamin~M
  Neale, and Mark~J Daly.
\newblock Clinical use of current polygenic risk scores may exacerbate health
  disparities.
\newblock {\em Nature genetics}, 51(4):584--591, 2019.

\bibitem{wang2023global}
Ying Wang, Shinichi Namba, Esteban Lopera, Sini Kerminen, Kristin Tsuo, Kristi
  L{\"a}ll, Masahiro Kanai, Wei Zhou, Kuan-Han Wu, Marie-Julie Fav{\'e}, et~al.
\newblock Global biobank analyses provide lessons for developing polygenic risk
  scores across diverse cohorts.
\newblock {\em Cell Genomics}, 3(1), 2023.

\bibitem{ge2019polygenic}
Tian Ge, Chia-Yen Chen, Yang Ni, Yen-Chen~Anne Feng, and Jordan~W Smoller.
\newblock Polygenic prediction via bayesian regression and continuous shrinkage
  priors.
\newblock {\em Nature communications}, 10(1):1776, 2019.

\bibitem{reyes2023fairprs}
Diego~Machado Reyes, Aritra Bose, Ehud Karavani, and Laxmi Parida.
\newblock Fairprs: adjusting for admixed populations in polygenic risk scores
  using invariant risk minimization.
\newblock In {\em Pacific Symposium on Biocomputing. Pacific Symposium on
  Biocomputing}, volume~28, page 198. NIH Public Access, 2023.

\bibitem{platt2024ai}
Daniel~E Platt, Aldo Guzm{\'a}n-S{\'a}enz, Aritra Bose, Subrata Saha, Filippo
  Utro, and Laxmi Parida.
\newblock Ai-enabled evaluation of genome-wide association relevance and
  polygenic risk score prediction in alzheimer's disease.
\newblock {\em Iscience}, 27(3), 2024.

\bibitem{wang2022application}
Yizhuo Wang, Bing~Z Carter, Ziyi Li, and Xuelin Huang.
\newblock Application of machine learning methods in clinical trials for
  precision medicine.
\newblock {\em JAMIA open}, 5(1):ooab107, 2022.

\bibitem{basu2023quantumenabled}
Saugata Basu, Jannis Born, Aritra Bose, Sara Capponi, Dimitra Chalkia,
  Timothy~A Chan, Hakan Doga, Frederik~F. Flother, Gad Getz, Mark Goldsmith,
  Tanvi Gujarati, Aldo Guzman-Saenz, Dimitrios Iliopoulos, Gavin~O. Jones,
  Stefan Knecht, Dhiraj Madan, Sabrina Maniscalco, Nicola Mariella, Joseph~A.
  Morrone, Khadijeh Najafi, Pushpak Pati, Daniel Platt, Maria~Anna Rapsomaniki,
  Anupama Ray, Kahn Rhrissorrakrai, Omar Shehab, Ivano Tavernelli, Meltem
  Tolunay, Filippo Utro, Stefan Woerner, Sergiy Zhuk, Jeannette~M. Garcia, and
  Laxmi Parida.
\newblock Towards quantum-enabled cell-centric therapeutics, 2023.

\bibitem{barkoutsos2020improving_QO}
Panagiotis~Kl Barkoutsos, Giacomo Nannicini, Anton Robert, Ivano Tavernelli,
  and Stefan Woerner.
\newblock Improving variational quantum optimization using cvar.
\newblock {\em Quantum}, 4:256, 2020.

\bibitem{moll2018quantum_QO_var}
Nikolaj Moll, Panagiotis Barkoutsos, Lev~S Bishop, Jerry~M Chow, Andrew Cross,
  Daniel~J Egger, Stefan Filipp, Andreas Fuhrer, Jay~M Gambetta, Marc Ganzhorn,
  et~al.
\newblock Quantum optimization using variational algorithms on near-term
  quantum devices.
\newblock {\em Quantum Science and Technology}, 3(3):030503, 2018.

\bibitem{symons2023practitioner}
Benjamin~CB Symons, David Galvin, Emre Sahin, Vassil Alexandrov, and Stefano
  Mensa.
\newblock A practitioner’s guide to quantum algorithms for optimisation
  problems.
\newblock {\em Journal of Physics A: Mathematical and Theoretical},
  56(45):453001, 2023.

\bibitem{cerezo2021variational}
Marco Cerezo, Andrew Arrasmith, Ryan Babbush, Simon~C Benjamin, Suguru Endo,
  Keisuke Fujii, Jarrod~R McClean, Kosuke Mitarai, Xiao Yuan, Lukasz Cincio,
  et~al.
\newblock Variational quantum algorithms.
\newblock {\em Nature Reviews Physics}, 3(9):625--644, 2021.

\bibitem{abbas2023quantum_optimization}
Amira Abbas, Andris Ambainis, Brandon Augustino, Andreas B{\"a}rtschi, Harry
  Buhrman, Carleton Coffrin, Giorgio Cortiana, Vedran Dunjko, Daniel~J Egger,
  Bruce~G Elmegreen, et~al.
\newblock Quantum optimization: Potential, challenges, and the path forward.
\newblock {\em arXiv preprint arXiv:2312.02279}, 2023.

\bibitem{dalzell2023quantum_ML}
Alexander~M. Dalzell, Sam McArdle, Mario Berta, Przemyslaw Bienias, Chi-Fang
  Chen, András Gilyén, Connor~T. Hann, Michael~J. Kastoryano, Emil~T.
  Khabiboulline, Aleksander Kubica, Grant Salton, Samson Wang, and Fernando G.
  S.~L. Brandão.
\newblock Quantum algorithms: A survey of applications and end-to-end
  complexities, 2023.

\bibitem{schuld2015introduction_QML}
Maria Schuld, Ilya Sinayskiy, and Francesco Petruccione.
\newblock An introduction to quantum machine learning.
\newblock {\em Contemporary Physics}, 56(2):172--185, 2015.

\bibitem{biamonte2017quantum_ML}
Jacob Biamonte, Peter Wittek, Nicola Pancotti, Patrick Rebentrost, Nathan
  Wiebe, and Seth Lloyd.
\newblock Quantum machine learning.
\newblock {\em Nature}, 549(7671):195--202, 2017.

\bibitem{cerezo2022challenges_QML}
M~Cerezo, Guillaume Verdon, Hsin-Yuan Huang, Lukasz Cincio, and Patrick~J
  Coles.
\newblock Challenges and opportunities in quantum machine learning.
\newblock {\em Nature Computational Science}, 2(9):567--576, 2022.

\bibitem{cong2019quantum_CNN}
Iris Cong, Soonwon Choi, and Mikhail~D Lukin.
\newblock Quantum convolutional neural networks.
\newblock {\em Nature Physics}, 15(12):1273--1278, 2019.

\bibitem{otterbach2017unsupervised}
Johannes~S Otterbach, Riccardo Manenti, Nasser Alidoust, A~Bestwick, M~Block,
  B~Bloom, S~Caldwell, N~Didier, E~Schuyler Fried, S~Hong, et~al.
\newblock Unsupervised machine learning on a hybrid quantum computer.
\newblock {\em arXiv preprint arXiv:1712.05771}, 2017.

\bibitem{atzori2019second}
Matteo Atzori and Roberta Sessoli.
\newblock The second quantum revolution: role and challenges of molecular
  chemistry.
\newblock {\em Journal of the American Chemical Society}, 141(29):11339--11352,
  2019.

\bibitem{kovyrshin2023nonadiabatic}
Arseny Kovyrshin, M\r{a}rten Skogh, Lars Tornberg, Anders Broo, Stefano Mensa,
  Emre Sahin, Benjamin~CB Symons, Jason Crain, and Ivano Tavernelli.
\newblock Nonadiabatic nuclear--electron dynamics: a quantum computing
  approach.
\newblock {\em The Journal of Physical Chemistry Letters}, 14(31):7065--7072,
  2023.

\bibitem{nykanen2023toward}
Anton Nyk\"anen, Aaron Miller, Walter Talarico, Stefan Knecht, Arseny
  Kovyrshin, M\r{a}rten Skogh, Lars Tornberg, Anders Broo, Stefano Mensa,
  Benjamin~CB Symons, et~al.
\newblock Toward accurate post-born--oppenheimer molecular simulations on
  quantum computers: An adaptive variational eigensolver with
  nuclear-electronic frozen natural orbitals.
\newblock {\em Journal of Chemical Theory and Computation}, 19(24):9269--9277,
  2023.

\bibitem{kovyrshin2023quantum}
Arseny Kovyrshin, Marten Skogh, Anders Broo, Stefano Mensa, Emre Sahin, Jason
  Crain, and Ivano Tavernelli.
\newblock A quantum computing implementation of nuclearelectronic orbital (neo)
  theory: Toward an exact pre-born--oppenheimer formulation of molecular
  quantum systems.
\newblock {\em The Journal of Chemical Physics}, 158(21), 2023.

\bibitem{kandala2017hardware}
Abhinav Kandala, Antonio Mezzacapo, Kristan Temme, Maika Takita, Markus Brink,
  Jerry~M Chow, and Jay~M Gambetta.
\newblock Hardware-efficient variational quantum eigensolver for small
  molecules and quantum magnets.
\newblock {\em nature}, 549(7671):242--246, 2017.

\bibitem{rice2021quantum}
Julia~E Rice, Tanvi~P Gujarati, Mario Motta, Tyler~Y Takeshita, Eunseok Lee,
  Joseph~A Latone, and Jeannette~M Garcia.
\newblock Quantum computation of dominant products in lithium--sulfur
  batteries.
\newblock {\em The Journal of Chemical Physics}, 154(13), 2021.

\bibitem{eddins2022doubling}
Andrew Eddins, Mario Motta, Tanvi~P Gujarati, Sergey Bravyi, Antonio Mezzacapo,
  Charles Hadfield, and Sarah Sheldon.
\newblock Doubling the size of quantum simulators by entanglement forging.
\newblock {\em PRX Quantum}, 3(1):010309, 2022.

\bibitem{barison2022quantum}
Stefano Barison, Davide~E Galli, and Mario Motta.
\newblock Quantum simulations of molecular systems with intrinsic atomic
  orbitals.
\newblock {\em Physical Review A}, 106(2):022404, 2022.

\bibitem{motta2023quantum}
Mario Motta, Gavin~O Jones, Julia~E Rice, Tanvi~P Gujarati, Rei Sakuma, Ieva
  Liepuoniute, Jeannette~M Garcia, and Yu-ya Ohnishi.
\newblock Quantum chemistry simulation of ground-and excited-state properties
  of the sulfonium cation on a superconducting quantum processor.
\newblock {\em Chemical Science}, 14(11):2915--2927, 2023.

\bibitem{liepuoniute2024simulation}
Ieva Liepuoniute, Mario Motta, Thaddeus Pellegrini, Julia~E Rice, Tanvi~P
  Gujarati, Sofia Gil, and Gavin~O Jones.
\newblock Simulation of a diels-alder reaction on a quantum computer.
\newblock {\em arXiv preprint arXiv:2403.08107}, 2024.

\bibitem{rossmannek2023quantum}
Max Rossmannek, Fabijan Pavosevic, Angel Rubio, and Ivano Tavernelli.
\newblock Quantum embedding method for the simulation of strongly correlated
  systems on quantum computers.
\newblock {\em The Journal of Physical Chemistry Letters}, 14(14):3491--3497,
  2023.

\bibitem{bauer2020quantum}
Bela Bauer, Sergey Bravyi, Mario Motta, and Garnet Kin-Lic Chan.
\newblock Quantum algorithms for quantum chemistry and quantum materials
  science.
\newblock {\em Chemical Reviews}, 120(22):12685--12717, 2020.

\bibitem{ducas2018crystals}
L{\'e}o Ducas, Eike Kiltz, Tancrede Lepoint, Vadim Lyubashevsky, Peter Schwabe,
  Gregor Seiler, and Damien Stehl{\'e}.
\newblock Crystals-dilithium: A lattice-based digital signature scheme.
\newblock {\em IACR Transactions on Cryptographic Hardware and Embedded
  Systems}, pages 238--268, 2018.

\bibitem{scholten2024assessing}
Travis~L. Scholten, Carl~J. Williams, Dustin Moody, Michele Mosca, William
  Hurley, William~J. Zeng, Matthias Troyer, and Jay~M. Gambetta.
\newblock Assessing the benefits and risks of quantum computers, 2024.

\bibitem{glover2018tutorial}
Fred Glover, Gary Kochenberger, and Yu~Du.
\newblock A tutorial on formulating and using qubo models.
\newblock {\em arXiv preprint arXiv:1811.11538}, 2018.

\bibitem{gambella2020multiblock}
Claudio Gambella and Andrea Simonetto.
\newblock Multiblock admm heuristics for mixed-binary optimization on classical
  and quantum computers.
\newblock {\em IEEE Transactions on Quantum Engineering}, 1:1--22, 2020.

\bibitem{mensa2023quantum}
Stefano Mensa, Emre Sahin, Francesco Tacchino, Panagiotis Kl~Barkoutsos, and
  Ivano Tavernelli.
\newblock Quantum machine learning framework for virtual screening in drug
  discovery: a prospective quantum advantage.
\newblock {\em Machine Learning: Science and Technology}, 4(1):015023, 2023.

\bibitem{doga2023perspective}
Hakan Doga, Bryan Raubenolt, Fabio Cumbo, Jayadev Joshi, Frank~P. DiFilippo,
  Jun Qin, Daniel Blankenberg, and Omar Shehab.
\newblock A perspective on protein structure prediction using quantum
  computers, 2023.

\bibitem{orus2019quantum}
Rom{\'a}n Or{\'u}s, Samuel Mugel, and Enrique Lizaso.
\newblock Quantum computing for finance: Overview and prospects.
\newblock {\em Reviews in Physics}, 4:100028, 2019.

\bibitem{sahin2024efficient}
M~Emre Sahin, Benjamin~CB Symons, Pushpak Pati, Fayyaz Minhas, Declan Millar,
  Maria Gabrani, Jan~Lukas Robertus, and Stefano Mensa.
\newblock Efficient parameter optimisation for quantum kernel alignment: A
  sub-sampling approach in variational training.
\newblock {\em arXiv preprint arXiv:2401.02879}, 2024.

\bibitem{10.1371/journal.pone.0283933}
Hideaki Kawaguchi.
\newblock Application of quantum computing to a linear non-gaussian acyclic
  model for novel medical knowledge discovery.
\newblock {\em PLOS ONE}, 18(4):e0283933, 2023.

\bibitem{flother2023state}
Frederik~F Fl{\"o}ther.
\newblock The state of quantum computing applications in health and medicine.
\newblock {\em Research Directions: Quantum Technologies}, 1:e10, 2023.

\bibitem{johansson2008physiologically}
Fredrik Johansson and Ronnie Paterson.
\newblock Physiologically based in silico models for the prediction of oral
  drug absorption.
\newblock {\em Drug Absorption Studies: In Situ, In Vitro and In Silico
  Models}, pages 486--509, 2008.

\bibitem{Nestorov2003}
Ivan Nestorov.
\newblock Whole body pharmacokinetic models.
\newblock {\em Clinical Pharmacokinetics}, 42(10):883--908, 2003.

\bibitem{khalil2011physiologically}
Feras Khalil and Stephanie L{\"a}er.
\newblock Physiologically based pharmacokinetic modeling: methodology,
  applications, and limitations with a focus on its role in pediatric drug
  development.
\newblock {\em Journal of Biomedicine and Biotechnology}, 2011, 2011.

\bibitem{krishnankutty2012data}
Binny Krishnankutty, Shantala Bellary, Naveen~BR Kumar, and Latha~S Moodahadu.
\newblock Data management in clinical research: an overview.
\newblock {\em Indian journal of pharmacology}, 44(2):168, 2012.

\bibitem{subbiah2023next}
Vivek Subbiah.
\newblock The next generation of evidence-based medicine.
\newblock {\em Nature medicine}, 29(1):49--58, 2023.

\bibitem{caro2022generalization}
Matthias~C Caro, Hsin-Yuan Huang, Marco Cerezo, Kunal Sharma, Andrew
  Sornborger, Lukasz Cincio, and Patrick~J Coles.
\newblock Generalization in quantum machine learning from few training data.
\newblock {\em Nature communications}, 13(1):4919, 2022.

\bibitem{gil2024understanding}
Elies Gil-Fuster, Jens Eisert, and Carlos Bravo-Prieto.
\newblock Understanding quantum machine learning also requires rethinking
  generalization.
\newblock {\em Nature Communications}, 15(1):1--12, 2024.

\bibitem{caro2021encoding}
Matthias~C Caro, Elies Gil-Fuster, Johannes~Jakob Meyer, Jens Eisert, and Ryan
  Sweke.
\newblock Encoding-dependent generalization bounds for parametrized quantum
  circuits.
\newblock {\em Quantum}, 5:582, 2021.

\bibitem{banchi2021generalization}
Leonardo Banchi, Jason Pereira, and Stefano Pirandola.
\newblock Generalization in quantum machine learning: A quantum information
  standpoint.
\newblock {\em PRX Quantum}, 2(4):040321, 2021.

\bibitem{huang2021power}
Hsin-Yuan Huang, Michael Broughton, Masoud Mohseni, Ryan Babbush, Sergio Boixo,
  Hartmut Neven, and Jarrod~R McClean.
\newblock Power of data in quantum machine learning.
\newblock {\em Nature communications}, 12(1):2631, 2021.

\bibitem{abbas2021power}
Amira Abbas, David Sutter, Christa Zoufal, Aur{\'e}lien Lucchi, Alessio
  Figalli, and Stefan Woerner.
\newblock The power of quantum neural networks.
\newblock {\em Nature Computational Science}, 1(6):403--409, 2021.

\bibitem{li2022concentration}
Guangxi Li, Ruilin Ye, Xuanqiang Zhao, and Xin Wang.
\newblock Concentration of data encoding in parameterized quantum circuits.
\newblock {\em Advances in Neural Information Processing Systems},
  35:19456--19469, 2022.

\bibitem{jones2015physiologically}
HM~Jones, Yuan Chen, Christopher Gibson, Tycho Heimbach, Neil Parrott,
  SA~Peters, Jan Snoeys, VV~Upreti, Ming Zheng, and SD~Hall.
\newblock Physiologically based pharmacokinetic modeling in drug discovery and
  development: a pharmaceutical industry perspective.
\newblock {\em Clinical Pharmacology \& Therapeutics}, 97(3):247--262, 2015.

\bibitem{ideker2001new}
Trey Ideker, Timothy Galitski, and Leroy Hood.
\newblock A new approach to decoding life: systems biology.
\newblock {\em Annual review of genomics and human genetics}, 2(1):343--372,
  2001.

\bibitem{claudino2007metabolomics}
Wederson~Marcos Claudino, Alessandro Quattrone, Laura Biganzoli, Marta Pestrin,
  Ivano Bertini, and Angelo Di~Leo.
\newblock Metabolomics: available results, current research projects in breast
  cancer, and future applications.
\newblock {\em Journal of clinical oncology}, 25(19):2840--2846, 2007.

\bibitem{peters2021physiologically}
Sheila~Annie Peters.
\newblock {\em Physiologically based pharmacokinetic (PBPK) modeling and
  simulations: principles, methods, and applications in the pharmaceutical
  industry}.
\newblock John Wiley \& Sons, 2021.

\bibitem{jones2013basic}
HM~Jones and Karen Rowland-Yeo.
\newblock Basic concepts in physiologically based pharmacokinetic modeling in
  drug discovery and development.
\newblock {\em CPT: pharmacometrics \& systems pharmacology}, 2(8):1--12, 2013.

\bibitem{maguire2009design}
TJ~Maguire, E~Novik, P~Chao, J~Barminko, Y~Nahmias, ML~Yarmush, and K-C Cheng.
\newblock Design and application of microfluidic systems for in vitro
  pharmacokinetic evaluation of drug candidates.
\newblock {\em Current drug metabolism}, 10(10):1192--1199, 2009.

\bibitem{schuhmacher2004high}
Joachim Schuhmacher, Christian Kohlsdorfer, Klaus B{\"u}hner, Tim
  Brandenburger, and Renate Kruk.
\newblock High-throughput determination of the free fraction of drugs strongly
  bound to plasma proteins.
\newblock {\em Journal of pharmaceutical sciences}, 93(4):816--830, 2004.

\bibitem{lloyd2020quantum}
Seth Lloyd, Giacomo De~Palma, Can Gokler, Bobak Kiani, Zi-Wen Liu, Milad
  Marvian, Felix Tennie, and Tim Palmer.
\newblock Quantum algorithm for nonlinear differential equations.
\newblock {\em arXiv preprint arXiv:2011.06571}, 2020.

\bibitem{liu2021efficient}
Jin-Peng Liu, Herman~{\O}ie Kolden, Hari~K Krovi, Nuno~F Loureiro, Konstantina
  Trivisa, and Andrew~M Childs.
\newblock Efficient quantum algorithm for dissipative nonlinear differential
  equations.
\newblock {\em Proceedings of the National Academy of Sciences},
  118(35):e2026805118, 2021.

\bibitem{krovi2023improved}
Hari Krovi.
\newblock Improved quantum algorithms for linear and nonlinear differential
  equations.
\newblock {\em Quantum}, 7:913, 2023.

\bibitem{cerezo2023does}
M~Cerezo, Martin Larocca, Diego Garc{\'\i}a-Mart{\'\i}n, NL~Diaz, Paolo
  Braccia, Enrico Fontana, Manuel~S Rudolph, Pablo Bermejo, Aroosa Ijaz,
  Supanut Thanasilp, et~al.
\newblock Does provable absence of barren plateaus imply classical
  simulability? or, why we need to rethink variational quantum computing.
\newblock {\em arXiv preprint arXiv:2312.09121}, 2023.

\bibitem{thanasilp2022exponential}
Supanut Thanasilp, Samson Wang, Marco Cerezo, and Zo{\"e} Holmes.
\newblock Exponential concentration and untrainability in quantum kernel
  methods.
\newblock {\em arXiv preprint arXiv:2208.11060}, 2022.

\bibitem{jaderberg2023let}
Ben Jaderberg, Antonio~A Gentile, Youssef~Achari Berrada, Elvira Shishenina,
  and Vincent~E Elfving.
\newblock Let quantum neural networks choose their own frequencies.
\newblock {\em arXiv preprint arXiv:2309.03279}, 2023.

\bibitem{paine2023quantum}
Annie~E Paine, Vincent~E Elfving, and Oleksandr Kyriienko.
\newblock Quantum kernel methods for solving regression problems and
  differential equations.
\newblock {\em Physical Review A}, 107(3):032428, 2023.

\bibitem{paine2023physics}
Annie~E Paine, Vincent~E Elfving, and Oleksandr Kyriienko.
\newblock Physics-informed quantum machine learning: Solving nonlinear
  differential equations in latent spaces without costly grid evaluations.
\newblock {\em arXiv preprint arXiv:2308.01827}, 2023.

\bibitem{rebentrost2018quantum}
Patrick Rebentrost and Seth Lloyd.
\newblock Quantum computational finance: quantum algorithm for portfolio
  optimization, 2018.

\bibitem{schuld2019quantum_kernel}
Maria Schuld and Nathan Killoran.
\newblock Quantum machine learning in feature hilbert spaces.
\newblock {\em Physical review letters}, 122(4):040504, 2019.

\bibitem{Kernel_method_EHR}
Zoran Krunic, Frederik~F. Flöther, George Seegan, Nathan~D. Earnest-Noble, and
  Omar Shehab.
\newblock Quantum kernels for real-world predictions based on electronic health
  records.
\newblock {\em IEEE Transactions on Quantum Engineering}, 3:1--11, 2022.

\bibitem{havlivcek2019supervised}
Vojt{\v{e}}ch Havl{\'\i}{\v{c}}ek, Antonio~D C{\'o}rcoles, Kristan Temme,
  Aram~W Harrow, Abhinav Kandala, Jerry~M Chow, and Jay~M Gambetta.
\newblock Supervised learning with quantum-enhanced feature spaces.
\newblock {\em Nature}, 567(7747):209--212, 2019.

\bibitem{article_EHR}
Danyal Maheshwari, Ubaid Ullah, Pablo Marulanda, Alain García-Olea, Ignacio
  Gonzalez, Jose Merodio, and Begoña Zapirain.
\newblock Quantum machine learning applied to electronic healthcare records for
  ischemic heart disease classification.
\newblock {\em Human-centric Computing and Information Sciences}, 13:17, 02
  2023.

\bibitem{Ghoabdi2023.08.09.552597}
Mohadeseh~Zarei Ghoabdi and Elaheh Afsaneh.
\newblock Quantum machine learning for untangling the real-world problem of
  cancers classification based on gene expressions.
\newblock {\em bioRxiv}, 2023.

\bibitem{mathur2022medical}
Natansh Mathur, Jonas Landman, Yun~Yvonna Li, Martin Strahm, Skander Kazdaghli,
  Anupam Prakash, and Iordanis Kerenidis.
\newblock Medical image classification via quantum neural networks, 2022.

\bibitem{Van_Heertum2015-rv}
Ronald Van~Heertum, Robert Scarimbolo, Robert Ford, Eli Berdougo, and J~Michael
  O'Neal.
\newblock Companion diagnostics and molecular imaging-enhanced approaches for
  oncology clinical trials.
\newblock {\em Drug Des. Devel. Ther.}, page 5215, September 2015.

\bibitem{Limeta2023-jn}
Angelo Limeta, Francesco Gatto, Markus~J Herrg{\aa}rd, Boyang Ji, and Jens
  Nielsen.
\newblock Leveraging high-resolution omics data for predicting responses and
  adverse events to immune checkpoint inhibitors.
\newblock {\em Comput Struct Biotechnol J}, 21:3912--3919, July 2023.

\bibitem{biom11111597}
EIena~I. Usova, Asiiat~S. Alieva, Alexey~N. Yakovlev, Madina~S. Alieva,
  Alexey~A. Prokhorikhin, Alexandra~O. Konradi, Evgeny~V. Shlyakhto, Paolo
  Magni, Alberico~L. Catapano, and Andrea Baragetti.
\newblock Integrative analysis of multi-omics and genetic approaches—a new
  level in atherosclerotic cardiovascular risk prediction.
\newblock {\em Biomolecules}, 11(11), 2021.

\bibitem{Huang2021}
Hsin-Yuan Huang, Michael Broughton, Masoud Mohseni, Ryan Babbush, Sergio Boixo,
  Hartmut Neven, and Jarrod~R. McClean.
\newblock Power of data in quantum machine learning.
\newblock {\em Nature Communications}, 12(1):2631, May 2021.

\bibitem{Havlíček2019}
Vojt{\v{e}}ch Havl{\'i}{\v{c}}ek, Antonio~D. C{\'o}rcoles, Kristan Temme,
  Aram~W. Harrow, Abhinav Kandala, Jerry~M. Chow, and Jay~M. Gambetta.
\newblock Supervised learning with quantum-enhanced feature spaces.
\newblock {\em Nature}, 567(7747):209--212, Mar 2019.

\bibitem{kirk2023emergent}
Jorja~J. Kirk, Matthew~D. Jackson, Daniel J.~M. King, Philip Intallura, and
  Mekena Metcalf.
\newblock Emergent order in classical data representations on ising spin
  models, 2023.

\bibitem{ragone2023representation}
Michael Ragone, Paolo Braccia, Quynh~T. Nguyen, Louis Schatzki, Patrick~J.
  Coles, Frederic Sauvage, Martin Larocca, and M.~Cerezo.
\newblock Representation theory for geometric quantum machine learning, 2023.

\bibitem{nguyen2022theory}
Quynh~T. Nguyen, Louis Schatzki, Paolo Braccia, Michael Ragone, Patrick~J.
  Coles, Frederic Sauvage, Martin Larocca, and M.~Cerezo.
\newblock Theory for equivariant quantum neural networks, 2022.

\bibitem{Kbler2021TheIB}
Jonas~M. K{\"u}bler, Simon Buchholz, and Bernhard Scholkopf.
\newblock The inductive bias of quantum kernels.
\newblock In {\em Neural Information Processing Systems}, 2021.

\bibitem{Schatzki2024}
L.~Schatzki, M.~Larocca, Q.T. Nguyen, et~al.
\newblock Theoretical guarantees for permutation-equivariant quantum neural
  networks.
\newblock {\em npj Quantum Information}, 10:12, 2024.

\bibitem{10.1007/978-3-030-00934-2_24}
Bastiaan~S. Veeling, Jasper Linmans, Jim Winkens, Taco Cohen, and Max Welling.
\newblock Rotation equivariant cnns for digital pathology.
\newblock In Alejandro~F. Frangi, Julia~A. Schnabel, Christos Davatzikos,
  Carlos Alberola-L{\'o}pez, and Gabor Fichtinger, editors, {\em Medical Image
  Computing and Computer Assisted Intervention -- MICCAI 2018}, pages 210--218,
  Cham, 2018. Springer International Publishing.

\bibitem{Kobak2019}
Dmitry Kobak and Philipp Berens.
\newblock The art of using t-sne for single-cell transcriptomics.
\newblock {\em Nature Communications}, 10(1):5416, Nov 2019.

\bibitem{Caro2022}
Matthias~C. Caro, Hsin-Yuan Huang, M.~Cerezo, Kunal Sharma, Andrew Sornborger,
  Lukasz Cincio, and Patrick~J. Coles.
\newblock Generalization in quantum machine learning from few training data.
\newblock {\em Nature Communications}, 13(1):4919, Aug 2022.

\bibitem{PhysRevLett.121.040502}
Seth Lloyd and Christian Weedbrook.
\newblock Quantum generative adversarial learning.
\newblock {\em Phys. Rev. Lett.}, 121:040502, Jul 2018.

\bibitem{Stein_2021}
Samuel~A. Stein, Betis Baheri, Daniel Chen, Ying Mao, Qiang Guan, Ang Li,
  Bo~Fang, and Shuai Xu.
\newblock Qugan: A quantum state fidelity based generative adversarial network.
\newblock In {\em 2021 IEEE International Conference on Quantum Computing and
  Engineering (QCE)}. IEEE, October 2021.

\bibitem{Zoufal2019}
Christa Zoufal, Aur{\'e}lien Lucchi, and Stefan Woerner.
\newblock Quantum generative adversarial networks for learning and loading
  random distributions.
\newblock {\em npj Quantum Information}, 5(1):103, Nov 2019.

\bibitem{kim2023evidence}
Youngseok Kim, Andrew Eddins, Sajant Anand, Ken~Xuan Wei, Ewout Van Den~Berg,
  Sami Rosenblatt, Hasan Nayfeh, Yantao Wu, Michael Zaletel, Kristan Temme,
  et~al.
\newblock Evidence for the utility of quantum computing before fault tolerance.
\newblock {\em Nature}, 618(7965):500--505, 2023.

\bibitem{bravyi2024high}
Sergey Bravyi, Andrew~W Cross, Jay~M Gambetta, Dmitri Maslov, Patrick Rall, and
  Theodore~J Yoder.
\newblock High-threshold and low-overhead fault-tolerant quantum memory.
\newblock {\em Nature}, 627(8005):778--782, 2024.

\end{thebibliography}

\end{document}